\documentclass[aps,prd,preprint,superscriptaddress,nofootinbib,amsfonts,amssymb,amsmath]{revtex4-1}
\usepackage{graphicx,bm,color}
\usepackage{wrapfig}
\usepackage{cancel} 

\begin{document}

\title{Cosmic History of  Chameleonic Dark Matter in $F(R)$ Gravity}

\author{Taishi Katsuragawa}
\email{taishi@th.phys.nagoya-u.ac.jp}
\altaffiliation[Current affiliation: ]{Institute of Astrophysics, Central China Normal University, Wuhan 430079, China}
\affiliation{Kobayashi-Maskawa Institute for the Origin of Particles and the Universe, Nagoya University, Nagoya 464-8602, Japan}

\author{Shinya Matsuzaki}
\email{synya@hken.phys.nagoya-u.ac.jp}
\affiliation{Department of Physics, Nagoya University, Nagoya 464-8602, Japan}
\affiliation{Institute for Advanced Research, Nagoya University, Nagoya 464-8602, Japan}

\begin{abstract}
We study the cosmic history of the scalaron in $F(R)$ gravity with constructing the time evolution of the cosmic environment and discuss the {\it chameleonic dark matter} based on the chameleon mechanism in the early and current Universe.
We then find that the scalaron can be a dark matter.
We also propose an interesting possibility that the $F(R)$ gravity can address the coincidence problem.
\end{abstract}

\maketitle

\section{Introduction}
\label{1}

\subsection{Motivation for the $F(R)$ gravity}
\label{1A}

Modified gravity theories have been investigated as a gravitational theory beyond the general relativity.
The necessity of such theories are motivated by many topics;
the modified gravity theory can be regarded as an effective theory and modeling of quantum gravity at an ultraviolet scale, while it would be responsible for the late-time cosmic acceleration at an infrared scale.
Although the general relativity is very simple and unique, 
so many modified gravity theories have been proposed, which include a lot of possible ways to extend the general relativity: 
to introduce the new terms, new field, or the new principle.
 
$F(R)$ gravity is one way of simplest extensions of the general relativity among the other modified gravity theories;
the gravitational action is replaced by the function of Ricci scalar, instead of the Einstein-Hilbert action.
One can show that the $F(R)$ gravity can be, through the mathematical procedure, written as the general relativity with an additional scalar field whose potential energy could cause the accelerated expansion of the Universe.
Thus, $F(R)$ gravity potentially brings us the solution to the two phenomena: 
the inflation and dark energy.
The famous model of $F(R)$ gravity for the inflation, the cosmic acceleration at the very early Universe, is called the $R^2$ inflation \cite{Starobinsky:1980te} which includes $R^2$ correction to the Einstein-Hilbert action.
This model predicts the slow-roll and large field inflation, 
which is consistent with recent observational data.
The $F(R)$ gravity models for the dark energy has also been proposed (for a recent review, see \cite{Nojiri:2017ncd}), and they can explain the late-time cosmic acceleration instead of the general relativity with a cosmological constant.
In these models, the dark energy is not explained by the constant vacuum energy, but the energy of dynamical scalar field.

\subsection{Scalar field as dark matter?}
\label{1B}

As we have discussed in the previous subsection, the scalar field plays an important role in the $F(R)$ gravity.
Hereafter, we call this scalar field as {\it scalaron} to distinguish it from the other scalar fields which are induced from the theories beyond the standard model of particle physics.
It is significant to note that
the scalaron field is not related to the extension of the standard model, 
but derived from the modification of gravitational theory. 
Then, the problem is how to assign the role of physics to the scalaron.
For example, the scalaron is identified as an inflaton in the $F(R)$ model for the inflation.
In the $F(R)$ model for the dark energy, the scalaron is regarded as a dynamical dark energy.

In our previous work \cite{Katsuragawa:2016yir}, we proposed the scenario in which the scalaron can be a new dark matter candidate.
Similar topics had been researched in many literatures 
\cite{Nojiri:2008nk,Nojiri:2008nt,Cembranos:2008gj,Choudhury:2015zlc}.
We postulated that 
the fluctuation of the scalaron field around the potential minimum can be regarded as dark matter, just like in the case of the axion field,
while the potential energy explains the dark energy with a usual manner in the viable $F(R)$ gravity for the dark energy.
This new dark matter candidate has a particular property, known as the chameleon mechanism;
the scalaron mass depends on the environment surrounding the scalaron field.
We investigated the couplings between the scalaron and the standard model particles and estimated the lifetime of the scalaron.
We then placed the limit on parameters in the Starobinsky model of $F(R)$ gravity 
and discussed the validity of the scenario in which the scalaron can be the dark matter candidate.
We found that the scalaron mass becomes too large,
and then, the parameter in the Starobinsky model is tightly constrained.
We finally found that such a parameter region is not relevant to the dark energy problem.

In this paper, we reconsider and refine our previous proposal with developed evaluation in the different $F(R)$ gravity model.
And we study again if the scalaron can be a dark matter candidate
with an identification that the oscillation of the scalaron gives us a particle picture. 
We study the cosmic history of the scalaron with constructing the time evolution of the cosmic environment,
and discuss the {\it chameleonic dark matter} based on the chameleon mechanism in the early and current Universe.

\section{$F(R)$ gravity and Scalaron}
\label{2}

\subsection{Action and Weyl transformation}
\label{2A}

We begin with the action of generic $F(R)$ gravity:
\begin{align}
S=\frac{1}{2\kappa^{2}} \int d^{4}x \sqrt{-g} F(R) + \int d^{4}x \sqrt{-g} \mathcal{L}_\mathrm{Matter} [g^{\mu \nu}, \Phi]
\label{action1}
\, ,
\end{align}
where $F(R)$ is a function of the Ricci scalar $R$ and $\kappa^{2} = 8\pi G = 1/M^{2}_\mathrm{pl}$.
$M_\mathrm{pl}$ is the (reduced) Planck mass 
$\sim 2 \times 10^{18} [\mathrm{GeV}]$.
$\mathcal{L}_\mathrm{Matter}$ denotes the Lagrangian for a matter field $\Phi$.

The variation with respect to the metric $g_{\mu \nu}$ leads to the equation of motion:
\begin{align}
F_{R}(R) R_{\mu \nu} - \frac{1}{2} F(R) g_{\mu \nu} + (g_{\mu \nu} \Box - \nabla_{\mu} \nabla_{\nu}) F_{R}(R) 
= \kappa^{2} T_{\mu \nu} (g^{\mu \nu}, \Phi)
\label{jordaneom1}
\, .
\end{align}
Here, $F_{R}(R)$ means the derivative of $F(R)$ with respect to $R$, $F_{R}(R) = \partial_{R} F(R)$, and the energy-momentum tensor $T_{\mu \nu}$ is given by
\begin{align}
T_{\mu \nu}(g^{\mu \nu}, \Phi) 
= \frac{-2}{\sqrt{-g}} \frac{\delta \left(\sqrt{-g} \mathcal{L}_\mathrm{Matter} (g^{\mu \nu}, \Phi) \right)}{\delta g^{\mu \nu}}
\, .
\end{align}
Note that the conservation of the energy-momentum tensor $\nabla_{\mu}T^{\mu \nu}=0$ is guaranteed by the generalized Bianchi identity. 
Furthermore, taking the trace of the equation of motion (\ref{jordaneom1}), we have 
\begin{align}
\Box F_{R}(R) 
= \frac{1}{3} \left[ 2 F(R) - R F_{R}(R)  + \kappa^{2} T (g^{\mu \nu}, \Phi) \right]
\label{jordaneom2}
\, ,
\end{align}
where we have momentarily used short-hand notations for the Ricci scalar as 
$R=R^{\mu}_{\ \mu}$ and for the trace of the energy-momentum tensor as 
$T = T^{\mu}_{\ \mu}$. 
Note that 
Eq.~(\ref{jordaneom2}) explicitly shows the deviation from the general relativity. 
Eq.~(\ref{jordaneom2}) also shows that the Ricci scalar $R$ becomes dynamical to be 
present as a new scalar degree of freedom in the   
$F(R)$ gravity:    
in the general relativity where $F(R)=R$, Eq.~(\ref{jordaneom2}) just leads to 
the trivial solution $R=-\kappa^{2}T$. 
Note also that if the equation of motion has a solution that $R_{\mu \nu} = \Lambda g_{\mu \nu}$, or $R=4\Lambda$, we obtain
\begin{align}
2 F(R) - R F_{R}(R) + \kappa^{2} T (g^{\mu \nu}, \Phi) = 0
\label{jordaneom3}
\, .
\end{align}
In the case of the positive $\Lambda$, 
we have the de Sitter solution in the $F(R)$ gravity.

We can 
look into the dynamics of the new scalar field via the Weyl transformation.
It is known that the $F(R)$ gravity is equivalent to the scalar-tensor theory via the Weyl transformation of the metric, which is the frame transformation from the Jordan frame $g_{\mu \nu}$ to the Einstein frame $\tilde{g}_{\mu \nu}$:
\begin{align}
g_{\mu \nu} \rightarrow \tilde{g}_{\mu \nu} 
= 
\mathrm{e}^{2 \sqrt{1/6} \kappa \varphi } g_{\mu \nu} 
\equiv  F_{R} (R) g_{\mu\nu}   
\label{Weyltrans}
\, .
\end{align}
Under the Weyl transformation, the original action Eq.~(\ref{action1}) is transformed as follows:
\begin{align}
S
=&
\frac{1}{2\kappa^{2}} \int d^{4}x \sqrt{-\tilde{g}} \tilde{R}  +
\int d^{4}x \sqrt{-\tilde{g}} \left[ - \frac{1}{2} \tilde{g}^{\mu \nu} (\partial_{\mu} \varphi) (\partial_{\nu} \varphi) - V(\varphi) \right]
\nonumber \\
& \qquad
+ \int d^{4}x \sqrt{-\tilde{g}} \, \mathrm{e}^{-4 \sqrt{1/6} \kappa \varphi } 
\mathcal{L}_\mathrm{Matter} \left[ g^{\mu \nu} = \mathrm{e}^{2 \sqrt{1/6} \kappa \varphi }  \tilde{g}^{\mu \nu}, \Phi \right]
\label{action2}
\, .
\end{align}
The $\varphi$ is the scalaron field and its potential part reads 
\begin{align}
V(\varphi) 
= \frac{1}{2\kappa^{2}} \frac{R F_{R}(R) - F(R)}{F^{2}_{R}(R)}
\label{scalaronpotential}
\, .
\end{align}
Note that 
through the Weyl transformation in Eq.~(\ref{Weyltrans}), 
the Ricci scalar $R$ is given as a function of the scalaron field $\varphi$, 
like $R = R(\varphi)$.

By the variation of the action in Eq.~(\ref{action2}) 
with respect to the Einstein frame metric $\tilde{g}_{\mu \nu}$, 
we obtain
\begin{align}
\tilde{R}_{\mu \nu} - \frac{1}{2}\tilde{R} \tilde{g}_{\mu \nu}
- \frac{1}{2} 
\left[ -\frac{1}{2}  \tilde{g}_{\mu \nu}  (\partial^{\lambda} \varphi) (\partial_{\lambda} \varphi) + (\partial_{\mu} \varphi) (\partial_{\nu} \varphi) - V(\varphi)  \tilde{g}_{\mu \nu} \right] 
= \kappa^{2} \tilde{T}_{\mu \nu} \left( \mathrm{e}^{2 \sqrt{1/6} \kappa \varphi }  \tilde{g}^{\mu \nu}, \Phi \right)
\, ,
\end{align}
where the energy-momentum tensor in the Einstein frame ($\tilde{T}_{\mu \nu}$) is 
related to that in the Jordan frame ($\tilde{T}_{\mu\nu}$) as follows: 
\begin{align}
\tilde{T}_{\mu \nu} \left( \mathrm{e}^{2 \sqrt{1/6} \kappa \varphi }  \tilde{g}^{\mu \nu}, \Phi \right) 
=& \frac{-2}{\sqrt{-\tilde{g}}} 
\frac{\delta \left(\sqrt{-\tilde{g}} \, \mathrm{e}^{-4 \sqrt{1/6} \kappa \varphi }  \mathcal{L}_\mathrm{Matter} \left( \mathrm{e}^{2 \sqrt{1/6} \kappa \varphi }  \tilde{g}^{\mu \nu}, \Phi \right) \right)}{\delta \tilde{g}^{\mu \nu}}
\nonumber \\
=&
\frac{-2}{ \mathrm{e}^{4 \sqrt{1/6} \kappa \varphi } \sqrt{-g}} 
\frac{\delta \left(\sqrt{-g} \mathcal{L}_\mathrm{Matter} \left( g^{\mu \nu}, \Phi \right) \right)}{ \mathrm{e}^{-2 \sqrt{1/6} \kappa \varphi } \delta \tilde{g}^{\mu \nu}}
\nonumber \\
=&
\mathrm{e}^{-2 \sqrt{1/6} \kappa \varphi } T_{\mu \nu}(g^{\mu \nu}, \Phi)
\, .
\end{align}

The variation with respect to the scalaron field $\varphi$ gives us the equation of motion 
for the scalaron field, 
\begin{align}
0
=&
\sqrt{-\tilde{g}} \left[ \tilde{\Box} \varphi - V_{, \varphi}(\varphi) \right] 
+ \frac{\delta}{\delta \varphi} \left( \sqrt{-\tilde{g}} \, \mathrm{e}^{-4 \sqrt{1/6} \kappa \varphi } 
\mathcal{L}_\mathrm{Matter} \left[ \mathrm{e}^{2 \sqrt{1/6} \kappa \varphi }  \tilde{g}^{\mu \nu}, \Phi \right] \right)
\nonumber \\
=&
\sqrt{-\tilde{g}} \left[ \tilde{\Box} \varphi - V_{, \varphi}(\varphi) \right] 
+ \frac{\delta}{\delta \varphi} \left( \sqrt{-g} \mathcal{L}_\mathrm{Matter} [g^{\mu \nu}, \Phi] \right)
\, \label{EOM:scalaron}.
\end{align}
Noting 
\begin{align}
\delta g^{\mu \nu} 
=& 
\frac{2\kappa}{\sqrt{6}} \mathrm{e}^{2\sqrt{1/6}\kappa \varphi} \delta \varphi 
\tilde{g}^{\mu \nu} 
= \frac{2\kappa}{\sqrt{6}} g^{\mu \nu} \delta \varphi
\,, \nonumber \\
\frac{\delta}{\delta \varphi} 
=&
\frac{2\kappa}{\sqrt{6}} g^{\mu \nu} \frac{\delta}{\delta g^{\mu \nu}}
\, ,
\end{align}
we rewrite Eq.~(\ref{EOM:scalaron}) as 
\begin{align}
\tilde{\Box} \varphi  =&
V_{, \varphi}(\varphi) 
+ \frac{\kappa}{\sqrt{6}} \mathrm{e}^{-4 \sqrt{1/6} \kappa \varphi }  T_\mu^\mu
\label{scalaroneom}
\, .
\end{align}
From Eq.~(\ref{scalaroneom}),
we define the effective potential of the scalaron as follows:
\begin{align}
V_\mathrm{eff}(\varphi) = V(\varphi) - \frac{1}{4} \mathrm{e}^{-4 \sqrt{1/6} \kappa \varphi }  T^{\mu}_{\ \mu}
\label{scalaroneffectivepotential1}
\, .
\end{align}

We should note that 
Eq.~(\ref{scalaroneom}) in the Einstein frame corresponds to Eq.~(\ref{jordaneom2}) in the Jordan frame. 
It is also remarkable that the effective potential of the scalaron includes the trace of energy-momentum tensor $T_\mu^\mu$. 
In other words, the matter distributions affect the potential structure of the scalaron; thus, the scalaron mass depends on the matter contribution.
This feature is related to so-called chameleon mechanism, which we will see later in detail.

\subsection{Effective potential of scalaron}
\label{2B}

In this subsection, we discuss the minimum of the scalaron effective potential and the scalaron mass with the matter effect $T^{\mu}_{\ \mu}$.
The first derivative of the scalaron effective potential is evaluated as 
\begin{align}
V(\varphi)_{\mathrm{eff}, \varphi} 
&= 
V(\varphi)_{, \varphi} + \frac{\kappa}{\sqrt{6}} \mathrm{e}^{-4 \sqrt{1/6} \kappa \varphi }  T^{\mu}_{\ \mu}
\nonumber \\
=&
\frac{1}{\sqrt{6} \kappa} \frac{2F(R) - R F_{R}(R)}{F^{2}_{R}(R)}  + \frac{\kappa}{\sqrt{6}} \frac{T^{\mu}_{\ \mu}}{F^{2}_{R}(R)}  
\nonumber \\
=&
\frac{1}{\sqrt{6} \kappa} \left( \frac{2F(R) - R F_{R}(R) + \kappa^{2}T^{\mu}_{\ \mu} }{F^{2}_{R}(R)} \right)
\label{scalaroneffectivepotential2}
\, .
\end{align}
The minimum of the potential at $\varphi = \varphi_{\min}$ should satisfy 
\begin{align}
V(\varphi_{\min})_{\mathrm{eff}, \varphi} = 0
\, .
\end{align}
Through the Weyl transformation $\mathrm{e}^{2 \sqrt{1/6} \kappa \varphi_{\min} } =  F_{R} (R_{\min}) $, 
the $\varphi_{\min}$ is determined as  
\begin{align}
2F(R_{\min}) - R_{\min} F_{R}(R_{\min})  + \kappa^{2} T^{\mu}_{\ \mu} = 0
\label{potentialminimum1}
\, .
\end{align}
%
Note that Eq.~(\ref{potentialminimum1}) gives the same form as in Eq.~(\ref{jordaneom3}), 
and hence the equation of motion in the Jordan frame certainly corresponds to that in the Einstein frame. 

Next, we evaluate the scalaron mass $m_{\varphi}$.
The (square of) scalaron mass is defined as the value of the second derivative of the effective potential at the minimum.
The second derivative of the effective potential is evaluated as follows: 
\begin{align}
V(\varphi)_{\mathrm{eff}, \varphi \varphi} 
&= 
V(\varphi)_{, \varphi \varphi}  - \frac{2\kappa^{2}}{3} \mathrm{e}^{-4 \sqrt{1/6} \kappa \varphi }  T^{\mu}_{\ \mu}
\nonumber \\
&=
\frac{1}{3F_{RR}(R)}  \left(1 + \frac{ R F_{RR}(R)}{F_{R}(R)} - \frac{4 F(R) F_{RR}(R)}{F^{2}_{R}(R)} \right)  
- \frac{2\kappa^{2}}{3}  \frac{T^{\mu}_{\ \mu}}{F^{2}_{R}(R)}
\nonumber \\
&=
\frac{1}{3F_{RR}(R)}  \left(1 + \frac{ R F_{RR}(R)}{F_{R}(R)} - \frac{ 2 \left( 2 F(R) + \kappa^{2}T^{\mu}_{\ \mu} \right) F_{RR}(R)}{F^{2}_{R}(R)} \right)  
\label{scalaroneffectivepotential3}
\, .
\end{align}
Substituting Eq.~(\ref{potentialminimum1}) into Eq,~(\ref{scalaroneffectivepotential3}), we obtain
\begin{align}
m^{2}_{\varphi}
&= 
V(\varphi_{\min})_{\mathrm{eff}, \varphi \varphi} 
\nonumber \\
&=
\frac{1}{3F_{RR}(R_{\min})}  \left(1 - \frac{ R_{\min} F_{RR}(R_{\min})}{F_{R}(R_{\min})} \right) 
\, .\label{mass:formula1}
\end{align}
Note that since the $\varphi_{\min}$ or $R_{\min}$ is determined by Eq.~(\ref{potentialminimum1}),  
the scalaron mass changes according to the trace of the energy-momentum tensor $T_\mu^\mu$.

\subsection{Chameleon mechanism}
\label{2C}

In the previous subsections, we have explicitly seen that the effective potential and the mass of the scalaron depend on the matter contribution in $T_\mu^\mu$. 
In this subsection, we discuss the effect of the matter distribution to the scalaron mass.
As an example, we consider the Starobinsky model for the late-time acceleration \cite{Starobinsky:2007hu},
\begin{align}
F(R) = R - \beta R_{c} \left[ 1 - \left( 1 +  \frac{R^2}{R^{2}_{c}} \right)^{-n} \right]
\label{starobinsky_action1}
\, .
\end{align}
The $R_{c}$ is taken to be a typical energy scale,  
where the gravitational action deviates from the Einstein-Hilbert action, 
and one expects that $ R_{c} \sim H^{2}_{0} \sim \Lambda$.
The index $n$ and the parameter $\beta$ are chosen to be positive constants.  
Note that this is not the $R^2$ (or well-known Starobinsky) inflation model.

Because the curvature $R$ should be on the same order as the energy-momentum 
tensor $\kappa^{2}T_\mu^\mu$, 
and 
larger than the dark energy scale $R_{c}$ 
when we consider the higher density of matter than the dark energy density. 
Therefore, 
we 
work in the large curvature limit $R>R_{c}$.  
In the large curvature limit, one can approximate Eq.~(\ref{starobinsky_action1}) as 
\begin{align}
F(R) 
&\approx 
R - \beta R_{c} + \beta R_{c} \left( \frac{R}{R_{c}} \right)^{-2n}
\label{starobinsky_action2} 
\, .
\end{align}
(Here one could identify the $\beta R_{c}$ as the cosmological constant, effectively.) 
Then, the minimum of the potential is determined as (see Eq.~(\ref{potentialminimum1})) 
\begin{align}
0
=&
2F(R) - R F_{R} (R) + \kappa^{2} T^{\mu}_{\ \mu}
\nonumber \\
&=
R - 2 \beta R_{c} + 2 (n+1) \beta R_{c} \left( \frac{R}{R_{c}} \right)^{-2n} + \kappa^{2} T^{\mu}_{\ \mu}
\label{potentialminimum2}
\, .
\end{align}
In the large curvature limit $R>R_{c}$, one finds
\begin{align}
R \approx - \kappa^{2} T^{\mu}_{\ \mu}
\, .
\end{align}
We also assume that the matter contribution is approximately expressed as the pressure-less dust, $T_\mu^\mu=-\rho$, where $\rho$ is the matter energy density.
Then, the scalaron mass in Eq.~(\ref{mass:formula1}) 
is evaluated in the large curvature limit as 
\begin{align}
m_{\varphi}^{2} 
=& \frac{1}{3F_{RR}(R)}  \left(1 - \frac{ R F_{RR}(R)}{F_{R}(R)} \right)
\nonumber \\
&=
\frac{R_{c}}{6n (2n+1) \beta} \left( \frac{R}{R_{c}} \right)^{2(n+1)}   
\left(1 - 
\frac{ 2n (2n+1) \beta }{\left( \frac{R}{R_{c}} \right)^{2n+1} - 2n \beta  }
\right)
 \notag \\ 
& \approx
\frac{R_{c}}{6n (2n+1) \beta} \left( \frac{\kappa^{2} \rho}{R_{c} }  \right)^{2(n+1)} 
\label{scalaronmass1}
\, .
\end{align}

We find that the scalaron mass is given by the monotonically increasing function of the energy density $\rho$, 
and thus, the scalaron becomes heavy in the high-density region of matter, for example, in the Solar System although it becomes light in the low-density region, for instance, inter-galactic region.
This feature is called the chameleon mechanism which is one of the screening mechanism in the modified gravity \cite{Khoury:2003aq}.
Thanks for the chameleon mechanism, the scalaron field is screened around the Solar System, which makes the $F(R)$ gravity relevant to the observations.
On the other hand, in the low energy density environment, that is, on the cosmological scale,
the scalaron field becomes dynamical.

\subsection{Interactions with standard-model particles}
\label{2D}

In the previous section, we have seen how the chameleon mechanism works in the $F(R)$ gravity.
Here, we discuss the couplings between the scalaron and matters in detail.
First, we recall the matter Lagrangian in the action of Eq.~(\ref{action2}), which is given by
\begin{align}
S_{\mathrm{Matter}} 
=& \int d^{4}x \sqrt{-\tilde{g}} \, \mathrm{e}^{-4\sqrt{1/6}\kappa \varphi(x)}
\mathcal{L} \left( \mathrm{e}^{2\sqrt{1/6}\kappa \varphi(x)}
\tilde{g}^{\mu \nu}, \Psi \right)
\label{mattercoupling1}
\, .
\end{align}
One can find that the scalaron interacts with the matters through the dilatonic couplings. 
In this subsection, we just summarize the scalaron couplings to the standard model particles 
derived in the previous work by the authors \cite{Katsuragawa:2016yir}.

We consider the fluctuation around the minimum of the potential at $\varphi_{\min}$
\begin{align}
\varphi \rightarrow \tilde{\varphi} = \varphi_{\min} + \varphi 
\, .
\end{align}
We hereafter assume that 
the fluctuation $\varphi $ gives the particle picture of the scalaron.
The exponential form of the scalaron field appearing in Eq.~(\ref{mattercoupling1}) 
is then expanded around the minimum as well: 
\begin{align}
\mathrm{e}^{Q \kappa \varphi (x)} 
\rightarrow 
\mathrm{e}^{Q \kappa \tilde{\varphi}(x)} 
=& 
\mathrm{e}^{Q \kappa \varphi_{\min} } \mathrm{e}^{Q \kappa \varphi(x)} 
\, ,
\end{align}
where $Q$ is an arbitrary constant.
Then the scalaron couplings to matter fields in Eq.~(\ref{mattercoupling1}) 
are modified just by the background value $\mathrm{e}^{Q \kappa \varphi_{\min} }$, 
which   
implying difference in measurements between the Einstein and Jordan frames; 
the dimensionful observables differ in each frame due to the Weyl transformation of metric.
If $\kappa \varphi_{\min}$ is small enough, 
the background $\mathrm{e}^{Q \kappa \varphi_{\min} }$ can be ignored. 
In the Starobinsky model, one then finds 
\begin{align}
\mathrm{e}^{2 \sqrt{1/6} \kappa \varphi } 
=& 
1 - 2n \beta \left( \frac{R}{R_{c}} \right)^{-(2n+1)} 
\,, \nonumber \\ 
\textrm{or equivalently} \qquad 
|\kappa \varphi| 
=& 
 \frac{\sqrt{6}}{2} \left| 
\ln \left( 1 -2n \beta \left( \frac{R}{R_{c}} \right)^{-(2n+1)}  \right) \right| 
\, . 
\end{align}
Thus the large curvature limit $R_{c} < R$ 
corresponds to the small background situation $|\kappa \varphi| \ll 1$.
In the large curvature limit, one can then 
expand the exponential form of the scalaron field 
as 
\begin{align}
\mathrm{e}^{Q \kappa \tilde{\varphi}(x)} 
\approx &
1 \cdot \left( 1 + Q \kappa \varphi + \mathcal{O}(\kappa^{2}\varphi^{2}) \right)
\nonumber \\
=& 1 + Q \kappa \varphi + \mathcal{O}(\kappa^{2}\varphi^{2}) 
\label{mattercoupling2}
\, .
\end{align}


Substituting the expansion form in Eq.~(\ref{mattercoupling2}) into the matter Lagrangian in Eq.~(\ref{mattercoupling1}),
we can obtain the couplings between the scalaron and matter fields. 
Below we just list the results obtained in the previous work \cite{Katsuragawa:2016yir}
for each species of particles: 
\begin{itemize}
\item For massless vector field $V$=$A$(photon) and $G$(gluon)
\begin{align}
\mathcal{L}
=
- \frac{3 g^2_V}{8(4\pi)^2} \left( \frac{3}{2} \sqrt{\frac{1}{6}} \kappa \varphi \right) \, 
F_{\mu\nu}^{2}(V)
+ \mathcal{O}(\kappa^{2}\varphi^{2})
\label{Lag:FF}
\end{align}
where $F_{\mu \nu}(A) = \partial_{\mu} A_{\nu} - \partial_{\nu} A_{\mu}$ with 
the electromagnetic coupling $g_A=e$ 
and $F_{\mu\nu}(G)$ denotes the gluon field strength for $V=G$   
with the QCD coupling $g_G=g_s$. 
\item For massive vector field (weak bosons) with the mass $m_V$
\begin{align}
\mathcal{L}
=&
\frac{2\kappa \varphi }{\sqrt{6}} \cdot \frac{1}{2} m^{2}_{V} \tilde{g}^{\mu \nu} A_{\mu} A_{\nu} 
+ \mathcal{O}(\kappa^{2}\varphi^{2})
\,. \label{Lag:AA}
\end{align}
\item For massive fermion field (quarks and leptons) with the mass $m_F$
\begin{align}
\mathcal{L}
=&
\frac{\kappa \varphi }{\sqrt{6}}  \cdot m_{F} \bar{\psi^{\prime}} \psi^{\prime}  
+ \mathcal{O}(\kappa^{2}\varphi^{2}) 
\,, 
\label{Lag:psi}
\end{align}
where the fermion field $\psi$ is redefined as 
$ \psi \rightarrow \psi^{\prime} = \mathrm{e}^{-3/2 \sqrt{1/6} \kappa \varphi} \psi $.
\end{itemize}

\subsection{Limit on scalaron mass}
\label{2E}

Given 
the couplings between the scalaron and the standard model particles
which we have listed in the previous section, 
one can calculate the decay processes of the scalaron to the standard model particle pair.
In the previous work \cite{Katsuragawa:2016yir}, 
we evaluated the decay width and the lifetime of the scalaron,
assuming that the scalaron is surrounded by the perfect fluid composed of the elementary particles, 
which would be realized in the early Universe 
before the QCD phase transition after the electroweak phase transition.
By considering all decay processes to the standard model particles,
we obtained the limit on the scalaron mass $m_{\varphi} < 0.23 [\mathrm{GeV}]$.
However, 
this estimation was not sufficient 
because the mass of scalaron should change according to the time-evolution of environment
due to the chameleon mechanism.

\begin{figure}[htbp]
\centering
\includegraphics[width=0.6\textwidth]{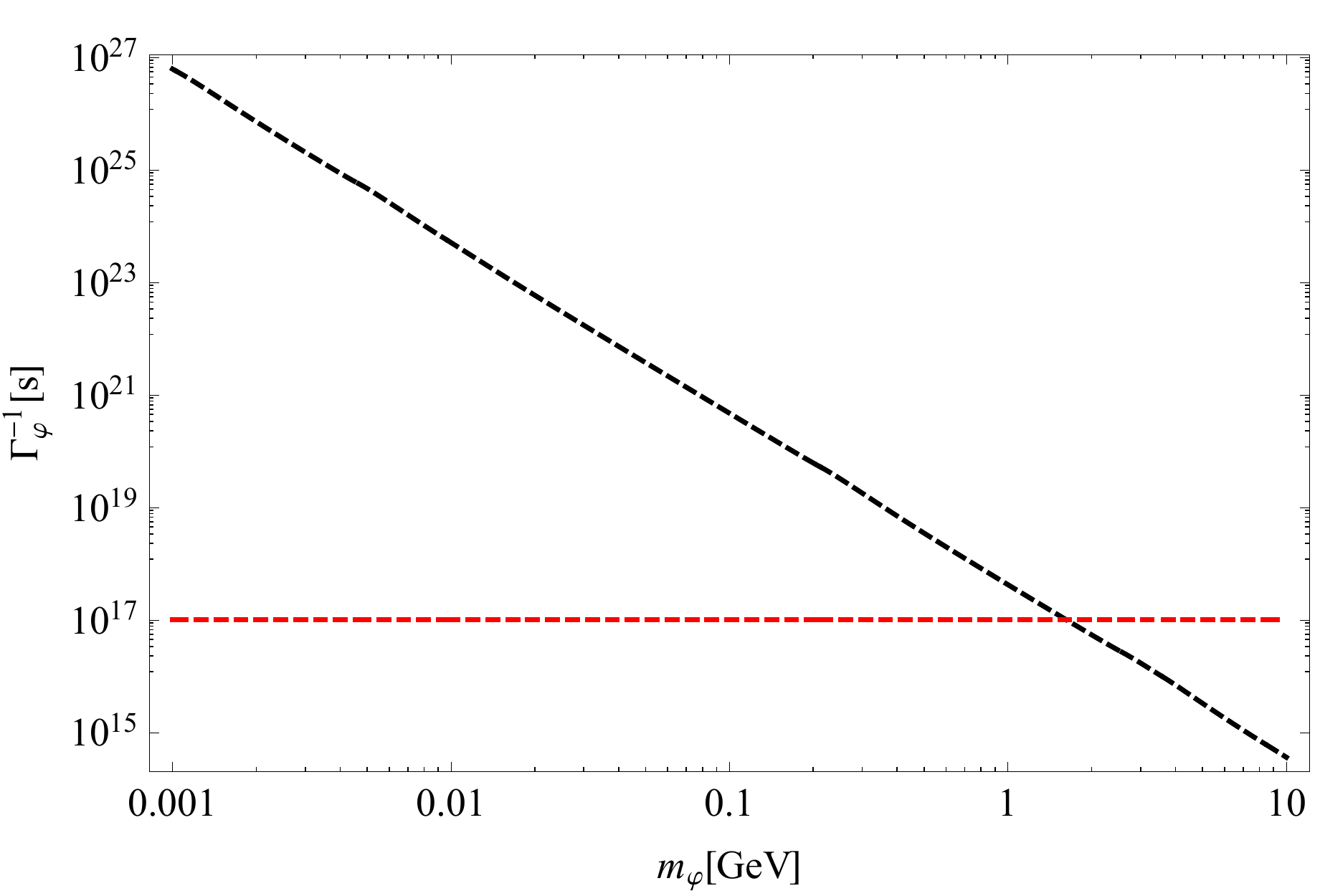}
\caption{The black dashed line shows the decay width of the scalaron to diphoton. 
The red dashed line denotes the age of the Universe.}
\label{scalaronlifetime}
\end{figure}

In this subsection,
we refine our result of the limit on the scalaron mass
with taking into account the cosmic history.
In the evaluation of the limit on the scalaron mass in \cite{Katsuragawa:2016yir}, 
we assumed that the lifetime of the scalaron should be longer than the age of Universe.
This argument is reasonable, except the environment assumed in the previous work \cite{Katsuragawa:2016yir} keeps only in very short time.
The decay processes to the massless particle pair would continue until the late time because the scalaron mass becomes small due to decreasing of 
the averaged energy density of the bulk. 
If the scalaron stull survives in the present Universe, 
the possible decay process would thus be dominated by 
the decay to the diphoton. 
Considering only the decay of the scalaron to two photons evaluated from 
the coupling form in Eq.(\ref{Lag:FF}),
we evaluate the lifetime of the scalaron as a function of the mass as depicted in 
 Fig.~\ref{scalaronlifetime} (For detailed evaluation, see~\cite{Katsuragawa:2016yir}). 
From this analysis, we find the more realistic upper bound for the scalaron mass:
\begin{align}
m_{\varphi} < \mathcal{O}(1)[\mathrm{GeV}]
\label{scalaronmassupperbound}
\, .
\end{align}
Our new result is comparable to the previous one.
Therefore, we can conclude that 
the scalaron mass at present should be smaller than or order of $1 [\mathrm{GeV}]$.


\section{Chameleonic dark matter}
\label{3}

\subsection{Scalaron in other model}
\label{3A}

In the last part of the previous section, 
we discussed the upper bound for the scalaron mass derived from 
the constraint on the lifetime.
In the previous work \cite{Katsuragawa:2016yir}, 
we found that the parameter $\beta$ in the Starobinsky model with $n=1$ should be extremely small $\beta \ll \mathcal{O}(1)$.
It is irrelevant to the solution to the dark energy problem $\beta > \mathcal{O}(1)$, which is also true for the presently refined result 
for the scalaron mass in Eq.~(\ref{scalaronmassupperbound}). 
This is because the scalaron becomes too heavy due to the chameleon mechanism;
Eq.~(\ref{scalaronmass1}) tells us that
the scalaron mass is not upper-bounded and can be heavier than the Planck mass if we consider a certain value of the energy density for matters.
Thus, the chameleon mechanism itself matters although it is the essential key in this scenario.

In this section, we re-establish our scenario to change the model of $F(R)$ gravity.
We will see that the extremely large scalaron mass is just an artifact in the Starobinsky model, which is possibly interpreted as a consequence of the singularity problem in the $F(R)$ gravity.
It has been suggested that one of the solutions to the singularity problem is to add the higher curvature term correction. 
Then, we will consider the Starobinsky model with a correction of $R^{2}$ term
and see that the scalaron mass can be lowered and controlled by the coefficients of $R^{2}$ term.
Here, we note that the $R^{2}$ term does not necessarily 
play the role of 
the $R^{2}$ inflation.

\subsection{Singularity problem}
\label{3B}

\begin{figure}[htbp]
\centering
\includegraphics[width=0.6\textwidth]{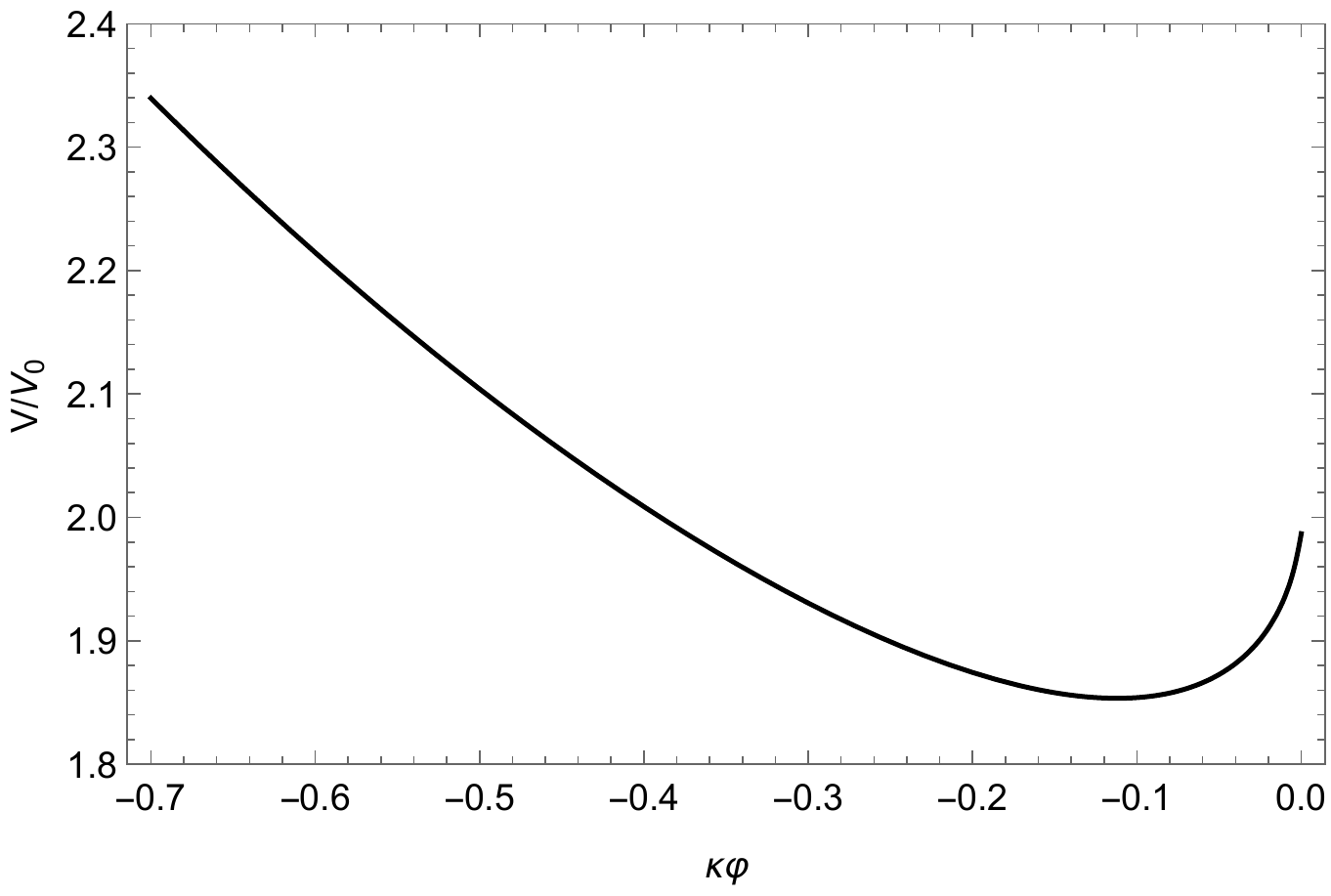}
\caption{The original potential of the scalaron $V$ in the Starobinsky model is plotted as a function of $\kappa \varphi$,
where the potential $V$ is normalized by the factor 
$V_{0} = \frac{R_{c}}{2\kappa^{2}}$.
Parameters are chosen as $n=1$, and $\beta=2$.}
\label{scalaronpotential_fig1}
\end{figure}

To understand the singularity problem, 
we first consider the scalaron potential without the matter contributions. 
From Eqs.~(\ref{starobinsky_action1}) and (\ref{scalaronpotential}),
we obtain the scalaron potential in the Starobinsky model as a function of the Ricci curvature $R$.
By using the Weyl transformation in Eq.~(\ref{Weyltrans}), which gives the relation between the Ricci curvature $R$ and scalaron field $\varphi$,
we obtain the scalaron potential as a function of the scalaron field.
The scalaron potential in the Starobinsky model is drawn in Fig.~\ref{scalaronpotential_fig1},
where the potential $V(\varphi)$ is normalized by the factor 
$V_{0} = \frac{R_{c}}{2\kappa^{2}}$,
and the parameters are chosen as $n=1$ and $\beta=2$.

One can find that the potential has a minimum at $\kappa \varphi_{\min} \sim 0.1$,
and then, the minimum is evaluated as $V(\varphi_{\min}) \sim \frac{2R_{c}}{2\kappa^{2}}$.
Recall that we expect $R_{c} \sim \Lambda$.  
Hence the minimum of the scalaron potential produces the effective cosmological constant 
$V(\varphi_{\min}) \sim \frac{2\Lambda}{2\kappa^{2}}$.
Note also that $\varphi=0$ corresponds to $R = \infty$;
the Weyl transformation gives the relation between $R$ and $\varphi$ as follows, 
\begin{align}
\mathrm{e}^{2 \sqrt{1/6} \kappa \varphi } 
=  1 - 2n \beta \left( \frac{R}{R_{c}} \right)^{-(2n+1)} 
\rightarrow 1
\ \mbox{when} \ 
R \rightarrow \infty
\, .
\end{align}
Here, we used Eq.~(\ref{starobinsky_action2}) because we work in the large-curvature limit.

\begin{figure}[htbp]
\centering
\includegraphics[width=0.6\textwidth]{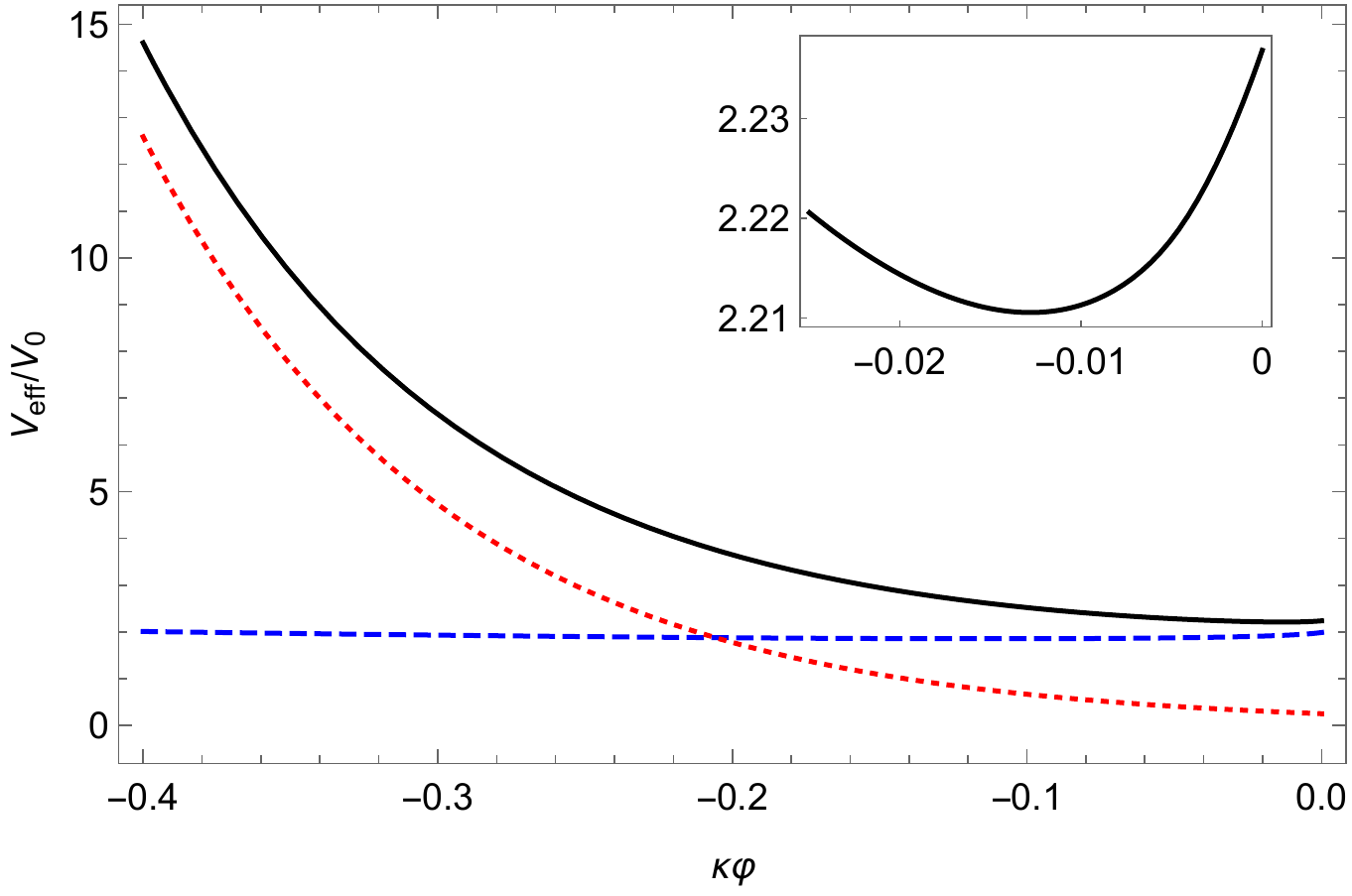}
\caption{The same as Fig.~\ref{scalaronpotential_fig1}  for the effective potential of the scalaron $V_{\mathrm{eff}}$ (black curves).
The blue dashed line denotes the original potential, and the red dotted line stands for the matter contribution with $-T^{\mu}_{\ \mu} \sim  \Lambda/2\kappa^{2} \sim \rho_{\Lambda}$.}
\label{scalaroneffectivepotential_fig1}
\end{figure}

Next, we add the contribution from the trace of energy-momentum tensor,
$- \frac{1}{4} \mathrm{e}^{-4 \sqrt{1/6} \kappa \varphi }  T^{\mu}_{\ \mu}$,
to discuss the effective potential of the scalaron as defined in Eq~(\ref{scalaroneffectivepotential1}). 
A sample plot of the effective potential is drawn in Fig.~\ref{scalaroneffectivepotential_fig1}.
By taking account the matter contribution, for the positive energy density of matter, we find that the potential is lifted up, 
and the $\varphi_{\min}$ becomes closer to zero.
Here, one can see that
the minimum of the effective potential is so shallow 
and the scalaron field at the minimum $\varphi = \varphi_{\min}$ can smoothly go to zero, 
$\varphi=0$.
This implies the drastic consequence 
that the curvature singularity is easily accessible because $\varphi = 0$ corresponds to $R=\infty$.
This is called the singularity problem \cite{Frolov:2008uf}, and it was suggested that 
viable $F(R)$ gravity models with the infrared modification generally suffer from this problem.

We now consider the singularity problem from the viewpoint of scalaron mass.
The matter effect becomes more eminent 
for the larger energy density in the effective potential,
and the minimum of the potential becomes shallower.
At the same time, the second derivative of the potential becomes exponentially large with respect to the energy density, which can be seen in the expression of the scalaron mass in Eq.~(\ref{scalaronmass1}).
We can, therefore, interpret the extremely heavy scalaron as a byproduct of the singularity problem,
and expect that the light, but not too light, scalaron in the high-density region would be realized if the singularity problem is resolved.

\subsection{Starobinsky model with $R^{2}$ correction}
\label{3C}
In the previous subsection, 
we have seen the singularity problem in $F(R)$ gravity in terms of the scalaron potential. 
In this subsection, we consider the prescription for the singularity problem.
A well-known way to this problem is to add the higher curvature term \cite{Nojiri:2008fk,Dev:2008rx,Bamba:2008ut,Kobayashi:2008wc,Capozziello:2009hc};
the problem arises in the large curvature region, thus, 
one naively expects that the singularity problem would be cured 
by improving the structure of the scalaron potential in the high energy regime.
We know that the quantum corrections of gravity are written as the higher power of the curvature tensor. 
Since we are studying the $F(R)$ gravity,
the relevant higher-curvature correction is given by the following form:
\begin{align}
F(R) = R - \beta R_{c} \left[ 1 - \left( 1 +  \frac{R^2}{R^{2}_{c}} \right)^{-n} \right] + \alpha R^{2}
\label{starobinsky_action3}
\, .
\end{align}
In the large curvature limit $R_{c}<R$, Eq.~(\ref{starobinsky_action3}) becomes
\begin{align}
F(R) 
\approx& R - \beta R_{c} \left[  1 - \left( \frac{R}{R_{c}} \right)^{-2n} \right]  + \alpha R^{2}
\nonumber \\
=& R - \beta R_{c} + \beta R_{c} \left( \frac{R}{R_{c}} \right)^{-2n}   + \alpha R^{2}
\label{starobinsky_action4}
\, . 
\end{align}
Then the relation between $R$ and $\varphi$ through the Weyl transformation and 
and the behavior in the large curvature limit are given as
\begin{align}
\mathrm{e}^{2 \sqrt{1/6} \kappa \varphi } 
=  1 - 2n \beta \left( \frac{R}{R_{c}} \right)^{-(2n+1)} + 2 \alpha R
\rightarrow \infty
\ \mbox{when} \ 
R \rightarrow \infty
\, . 
\end{align}
Compared with the original Starobinsky model in Eq.~(\ref{starobinsky_action1}),
we find that $\varphi$ can take the positive value and the limit $\varphi \to \infty$ is achieved  when $R  \to \infty$. 

\begin{figure}[htbp]
\centering
\includegraphics[width=0.6\textwidth]{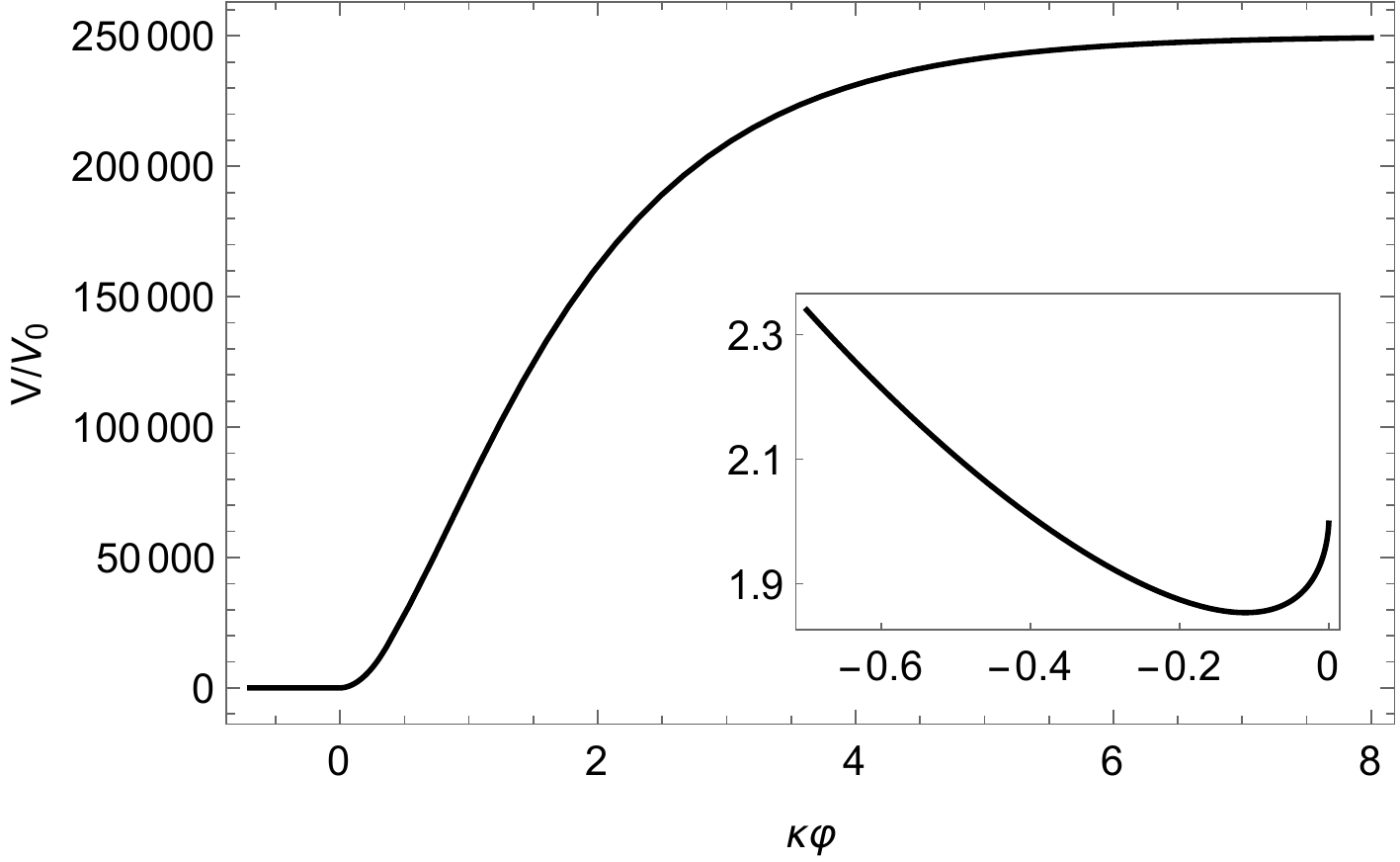}
\caption{The potential of the scalaron $V$ in the Starobinsky model with $R^2$ correction is plotted as a function of $\kappa \varphi$,
where the potential $V$ is normalized by the factor 
$V_{0} = \frac{R_{c}}{2\kappa^{2}}$.
Parameters are chosen as $n=1$, and $\beta=2$, and $\alpha = 10^{-6}/R_{c}$.}
\label{scalaronpotential_fig2}
\end{figure}

Next, we study the scalaron potential in the Starobinsky model with $R^{2}$ correction 
described by the $F(R)$ in Eq.~(\ref{starobinsky_action3}).
The scalaron potential $V$ 
is depicted in Fig.~\ref{scalaronpotential_fig2},
where the parameters are chosen as $n=1$, $\beta=2$, and $\alpha = 10^{-6}/R_{c}$.
The potential looks almost the same as the original one for the small curvature (negative $\varphi$) region, although it is modified for the large curvature (positive $\varphi$) region as we expected. 
Note that the potential in the large curvature limit is exactly the same as the one used in $R^{2}$ inflation scenario.

\begin{figure}[htbp]
\centering
\includegraphics[width=0.6\textwidth]{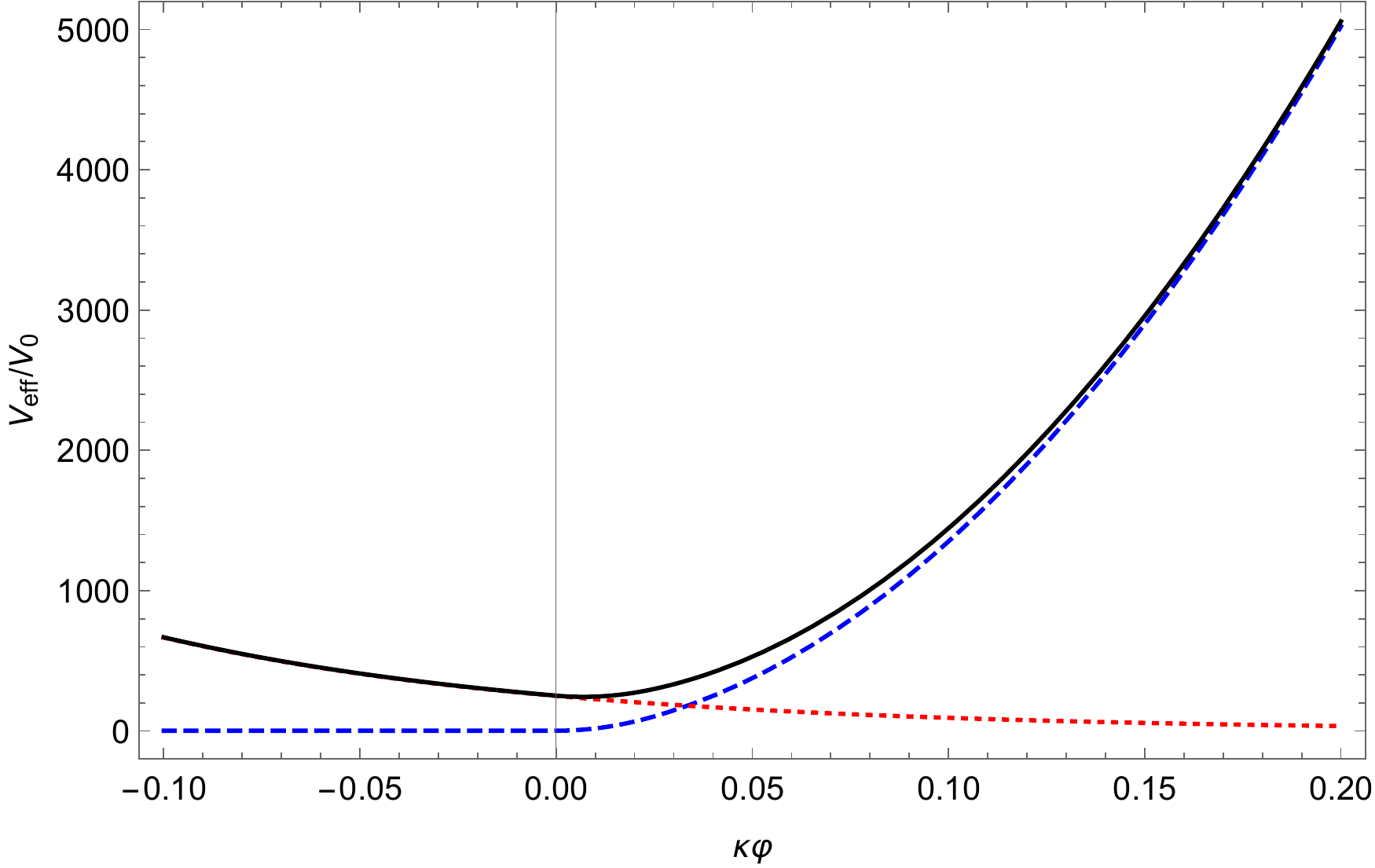}
\caption{The same as Fig.~\ref{scalaronpotential_fig2} for the effective potential of the scalaron $V_{\mathrm{eff}}$ (black curve).
The blue dashed line denotes the original potential, and the red dotted line stands for the matter contribution with $-T^{\mu}_{\ \mu} \sim 10^{3} \rho_{\Lambda} $.}
\label{scalaroneffectivepotential_fig2}
\end{figure}

We also examine the effective potential of the scalaron. 
The effective potential is drawn in Fig.~\ref{scalaroneffectivepotential_fig2}.
For the large energy density, the effective potential has the minimum in the large curvature region,
and prevents the minimum from being shallow.
Therefore, the singularity of the curvature is pushed away to infinity, 
which is not easily accessible. 
Thus the $R^{2}$ correction resolves the singularity problem in $F(R)$ gravity
while it does not make the infrared modification for the dark energy.

Finally, we discuss the scalaron mass in this model.
The minimum of the potential satisfies the stationary condition, 
\begin{align}
0
=&
2F(R) - R F_{R} (R) + \kappa^{2} T^{\mu}_{\ \mu}
\nonumber \\
=&
R - 2 \beta R_{c} + 2 (n+1) \beta R_{c} \left( \frac{R}{R_{c}} \right)^{-2n}  + \kappa^{2} T^{\mu}_{\ \mu}
\, .
\end{align}
Note that this equation is the same as Eq.~(\ref{potentialminimum2}):  
namely, 
the $R^{2}$ term does not contribute to the location of the minimum 
in terms of $R$.  
Therefore, in the large curvature limit $R_{c}<R$, one again finds  
\begin{align}
2F(R) - R F_{R} 
\approx
R = - \kappa^{2} T^{\mu}_{\ \mu}
\, .
\end{align}
Accordingly, the scalaron mass follows Eq.~(\ref{mass:formula1}), which is evaluated as 
\begin{align}
m_{\varphi}^{2} 
&= \frac{1}{3F_{RR}(R)}  \left(1 - \frac{ R F_{RR}(R)}{F_{R}(R)} \right)
\nonumber \\
&=
\frac{R_{c}}{6n (2n+1) \beta} 
\left[  \left( \frac{R}{R_{c}} \right)^{-2(n+1)}  + \frac{\alpha R_{c}}{n (2n+1) \beta}   \right]^{-1}
\frac{ 1 }{1  + 2\alpha R } 
\label{scalaronmass2}
\, .
\end{align}
In a moderately large curvature region where $R_{c}<R<1/\alpha$, 
the scalaron mass is expressed as follows:
\begin{align}
m_{\varphi}^{2} 
\approx
\frac{1}{6 \alpha}
\label{scalaronmass3}
\, .
\end{align}
This is the same as the well-known formula obtained in the $R^{2}$ inflation model. 
Thus the scalaron mass becomes constant and controlled by the parameter $\alpha$
in the large curvature region.
We also note that, 
If we worked in an extremely large curvature limit where $R> 1/\alpha$ (and $R>R_c)$, 
we could find 
\begin{align}
m_{\varphi}^{2} 
\approx 
\frac{1}{6 \alpha (1  + 2\alpha R ) } 
\label{scalaronmass4}
\, .
\end{align}
This implies that the scalaron mass decreases for such a very large curvature.situation. 
This is related to the fact that the very large energy density pushes 
the minimum of the potential up to be 
plateau.  

\begin{figure}[htbp]
\centering
\includegraphics[width=0.6\textwidth]{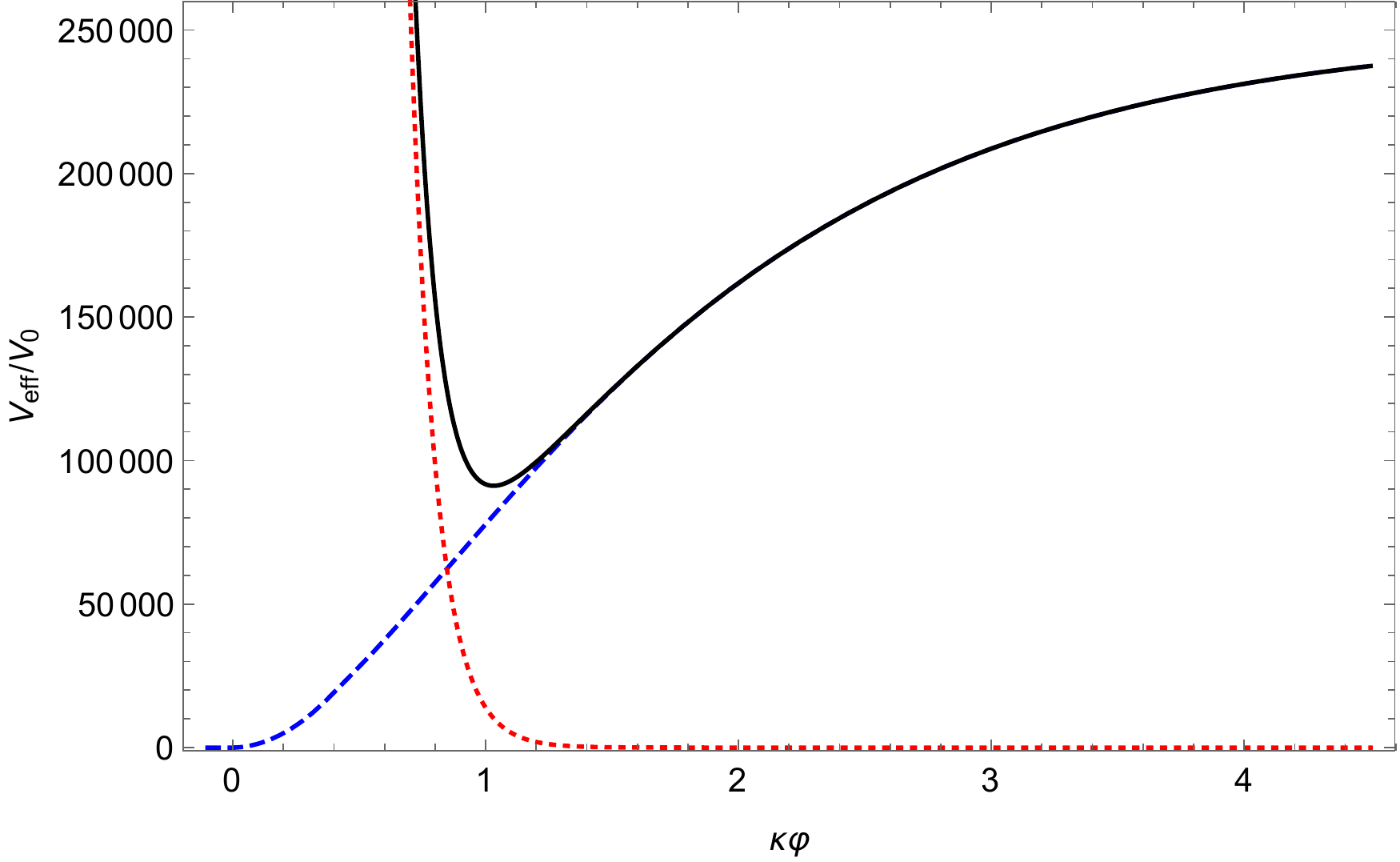}
\caption{The same as Fig.~\ref{scalaroneffectivepotential_fig2} for the effective potential of the scalaron $V_{\mathrm{eff}}$ (black curve).
The blue dashed line denotes the original potential, and the red dotted line stands for the matter contribution with $-T^{\mu}_{\ \mu} \sim  10^{9} \rho_{\Lambda} $.}
\label{scalaroneffectivepotential_fig10}
\end{figure}

We also comment on the value of  $\kappa \varphi$ at the minimum of the potential in the Starobinsky model with $R^{2}$ correction.
As we have discussed in Sec.~\ref{2D}, the background value $\mathrm{e}^{Q\kappa \varphi_{\min}}$ modifies the scalaron couplings to matter fields. 
In contrast to the original Starobinsky model, 
the location of the minimum $\kappa \varphi_{\min}$ becomes larger than ${\cal O}(1)$ 
if the trace of energy-momentum tensor ($-T^{\mu}_{\ \mu}$) is large enough 
in the Starobinsky model with $R^{2}$ correction. 
One can actually find $\kappa \varphi_{\min} \sim 1$ when $-T^{\mu}_{\ \mu} \sim 10^{9} \rho_{\Lambda} $, 
as shown in Fig.~\ref{scalaroneffectivepotential_fig10}.
Thus, the background-value effect for couplings to matter fields is of order one, 
$\mathrm{e}^{Q \kappa \varphi_{\min}} = \mathcal{O}(1)$
when $-T^{\mu}_{\ \mu} \lesssim 10^{9} \rho_{\Lambda} $. 
This result implies that the expansion form in Eq.~(\ref{mattercoupling2}) is approximately valid 
and the upper bound for the scalaron mass in Eq.~(\ref{scalaronmassupperbound}) in the original Starobinsky model is also applicable even when the $R^{2}$ correction term is included. 
We also note that the typical value of $T^{\mu}_{\ \mu}$ 
yielding $\kappa \varphi_{\min} \sim 1$ depends on the choice of the parameters.
Throughout this subsection, 
we have chosen the parameters as $n=1$, $\beta=2$, and $\alpha = 10^{-6}/R_{c}$. 
Actually, 
one can see that $\kappa \varphi_{\min}$ becomes smaller if smaller $\alpha$ is input.
As we will discuss in the next section, 
the reference value 
$\alpha = 10^{-6}/R_{c} \sim 10^{78} [\mathrm{GeV}^{-2}]$ is much 
larger than that favored by some experimental constraints, 
so the realistic $\kappa \varphi_{\min}$ is much smaller than $\mathcal{O}(1)$.
We can thus expect that the background-value effect can safely 
be neglected in the early Universe with the realistic parameter choice for $\alpha$.

\section{Cosmic History of Scalaron}
\label{4}

\subsection{Time-evolution of the scalaron mass}
\label{4A}

In the previous section, we studied the properties of the scalaron potential and mass in the Starobinsky model with $R^{2}$ correction. 
We have found that the scalaron does not become too massive
and the mass is given as $m^{2}_{\varphi} \propto \alpha^{-1}$ in the high density region.
In this subsection, we discuss the constraint on the parameters from the upper bound of the scalaron mass in Eq.~(\ref{scalaronmassupperbound}),
according to the time-evolution of the cosmological environment.
And, we also deduce a possible scenario of the cosmic history for the scalaron. 

First, we recall the relation between the scalaron mass and the chameleon mechanism.
In the Starobinsky model with $R^{2}$ correction, the scalaron mass scales as almost constant in the high-density region
and becomes smaller in the lower density region.
From the analysis in Sec.~\ref{2E}, we have obtained the limit on the scalaron mass $m_{\varphi}<\mathcal{O}(1)[\mathrm{GeV}]$ (see Eq.~(\ref{scalaronmassupperbound})).
One might think that 
this limit is converted to the constraint on the parameter $\alpha$, 
which would lead to $\alpha > \mathcal{O}(1)[\mathrm{GeV}^{-2}]$ 
from Eq.(\ref{scalaronmass3}). 
However, this argument does not make sense 
should not be present in the current Universe where the limit in Eq.~(\ref{scalaronmassupperbound}) has been placed. 
Thus we cannot place the constraint on the parameter $\alpha$ 
from the limit derived in the constant mass approximation as in Eq.~(\ref{scalaronmassupperbound}), 
which is valid only for the high dense region, i.e., the early Universe. 

Next, we consider the situation in the current Universe.
The energy density of matter decreases according to the cosmic evolution, and the scalaron becomes light at the late time.
Furthermore, if the environment is almost vacuum, the scalaron mass can be comparable to the dark energy scale.
The oscillation of scalaron around the potential minimum induces the very light particles, just like the case of axion,
and thus, the scalaron mass would naturally satisfy the upper bound 
$m_{\varphi}<\mathcal{O}(1)[\mathrm{GeV}]$ in Eq~.(\ref{scalaronmassupperbound}).
Therefore, we may deduce a natural scenario:  
the scalaron can be heavy in the early Universe but should be light, at least, in the current Universe.
If the scalaron is naturally light, we may obtain the parameter region relevant to the dark energy problem.

In the following subsections, we will demonstrate that this scenario is indeed realized 
by explicitly evaluating a part of the cosmic history of the scalaron:  
in principle, we can calculate everything, given the input of all cosmic history, 
to know the time evolution of the scalaron mass. 
It is, however, not so easy to follow all time-evolution of the Universe.  
In the following subsections, we will focus on the scalaron properties only in the early and current Universe.

\subsection{Environment in the early Universe}
\label{4B}

To closely study the scalaron potential, we need to evaluate the time-evolution of the energy-momentum tensor.
In the present analysis, we assume that the standard model particles only contribute to the energy-momentum tensor 
and the trace of energy-momentum tensor $T^{\mu}_{\ \mu} = - (\rho - 3p)$ can be approximated by the perfect fluid 
as done in \cite{Brax:2004qh}, 
where the pressure $p$ is not negligible. 
After the short calculation, we find the following expression (for detail, see the Appendix~\ref{A1});
\begin{align}
\rho - 3p 
= \frac{g T^{4}}{2 \pi^{2}} \cdot x^{2} \int^{\infty}_{0} dy \frac{y^{2}}{\sqrt{x^{2} + y^{2}}} \frac{1}{\mathrm{e}^{\sqrt{x^{2} + y^{2}}} \pm 1} 
\label{emttrace1}
\, ,
\end{align}
where the $+(-)$ sign is applied to fermions (bosons) and $g$ is the corresponding degrees of freedom. 
We have defined the variables $x\equiv\frac{m}{T}$ and $y \equiv \frac{p}{T}$, normalized by the temperature $T$. 

At the high temperature ($x \ll 1$), we obtain the following expression in the relativistic limit:
\begin{align}
\rho - 3p \approx \frac{g}{12}m^2T^{2}
\times 
\left\{
\begin{array}{l}
2\ \mbox{for bosons} \\
1\ \mbox{for fermions}
\end{array}
\right.
\label{emttrace2}
\, .
\end{align}
One can find that the temperature-dependence of the trace of $\rho - 3p$ is different from that of the energy density $\rho$;
$\rho - 3p \propto m^{2} T^{2}$ although $\rho \propto T^{4}$ in the relativistic limit.
Note that the pressure is also proportional to the temperature to the fourth, $p \propto T^{4}$.
Thus, the leading order terms ($\propto T^{4}$) for the energy density and pressure cancel each other, 
and the sub-leading term ($\propto m^{2} T^{2}$) remains. 
We also note that $\rho - 3p$ becomes exactly zero in the case of massless particles. 

At the low temperature ($x \gg 1$), we obtain the following expression in the non-relativistic limit:
\begin{align}
\rho - 3p 
\approx & 
\frac{g T^{4}}{2 \pi^{2}} \cdot x^{2}  
\sqrt{ \frac{\pi}{2} } x^{1/2} \mathrm{e}^{-x} \left(1  - \frac{3}{2}  x^{-1}  \right)
\nonumber \\
=&
m g \left( \frac{m T}{2 \pi} \right)^{3/2} \mathrm{e}^{-\frac{m}{T}}   - \frac{3}{2} g T \left( \frac{m T}{2 \pi} \right)^{3/2} \mathrm{e}^{-\frac{m}{T}} 
\label{emttrace3}
\, .
\end{align}
This expression can be rewritten as follows:
\begin{align}
\rho - 3p \approx \left( m - \frac{3}{2}  T \right)n
\,,
\end{align}
where $n$ is the number density defined as
\begin{align}
n \equiv g \left( \frac{m T}{2 \pi} \right)^{3/2} \mathrm{e}^{-\frac{m}{T}}
\, .
\end{align}

\subsection{Scalaron mass in the early Universe}
\label{4C}

\begin{figure}[htbp]
\centering
\includegraphics[width=0.6\textwidth]{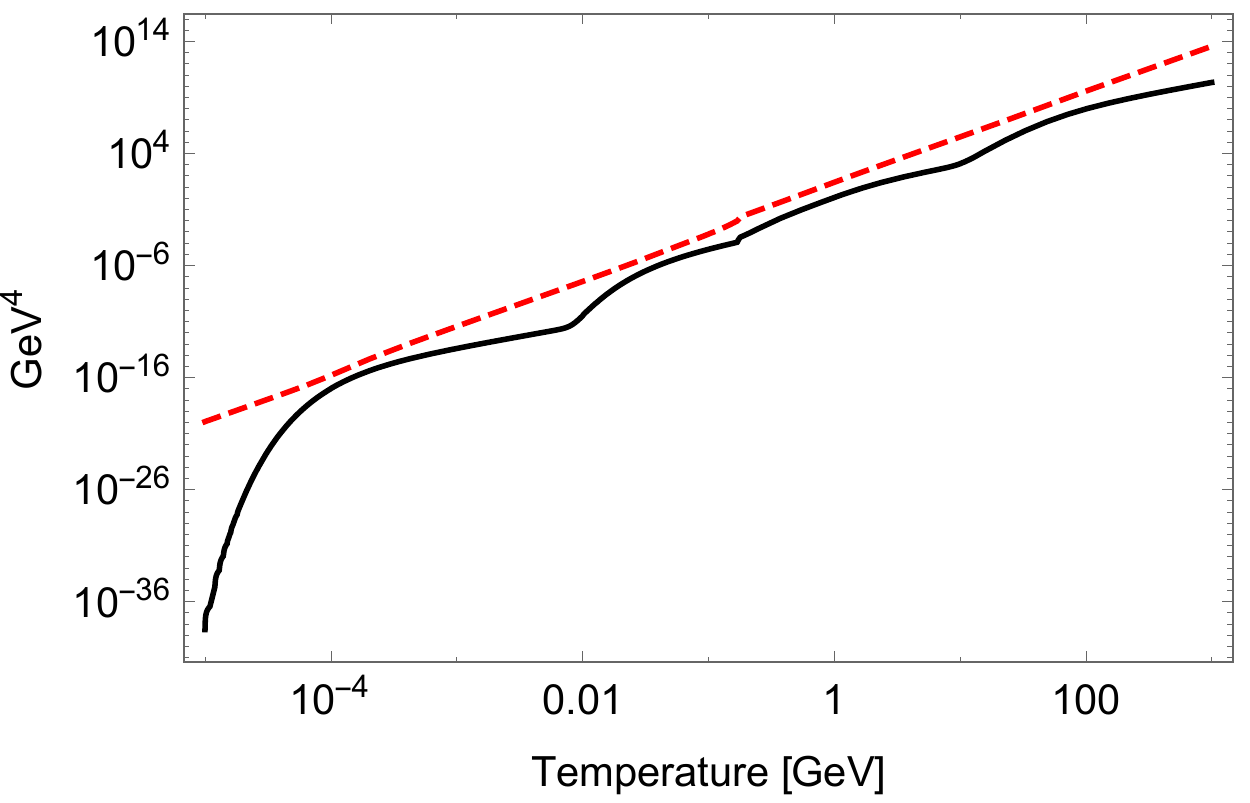}
\caption{The black curve shows the trace of energy-momentum tensor ($\rho - 3p$), and the red dashed curve corresponds to the energy density ($\rho$).}
\label{scalaronmasshistory_fig1}
\end{figure}

In the previous subsection, we derived the analytic formulae for the trace of the energy-momentum tensor for single species. 
In this subsection, we sum up all the contributions from the standard model particles 
including hadrons. 
The numerical result is shown in Fig.~\ref{scalaronmasshistory_fig1}.
Here, we have assumed that 
the critical temperature at the QCD phase transition is $T=170 [\mathrm{MeV}]$, 
and have used the approximation formula for the energy density $\rho$,  
\begin{align}
\rho = \frac{\pi^{2}}{30} g_{*}(T) T^{4}
\, ,
\end{align}
with $g_{*}(T)$, the effective degrees of freedom, taken from \cite{Husdal:2016haj}.  
Note that we can ignore the massless particles, photon and gluon, in the calculation of the trace of energy-momentum tensor because the massless particles do not contribute although they contribute to the energy density. 
We also assumed to ignore the contribution from neutrinos because they are almost massless. 
The input parameters for the standard model particles are summarized in the Table.~\ref{particlecontents}.
\begin{table}[htbp]
\centering
\begin{tabular}{lcrccc} \hline \hline
Type    &     &Mass  &D.O.F. &Before Phase Trans. &After Phase Trans. \\ \hline \hline
quarks    &$t$    &173[GeV] &12  &$\checkmark$   & \\ 
      &$b$    &4[GeV]  &12  &$\checkmark$   & \\ 
      &$c$    &1.2[GeV] &12  &$\checkmark$   & \\ 
      &$s$    &105[MeV] &12  &$\checkmark$   & \\ 
      &$d$    &5[MeV]  &12  &$\checkmark$   & \\ 
      &$u$    &2[MeV]  &12  &$\checkmark$   & \\ \hline
gluon    &$g$    &0    &16  &$\checkmark$   & \\ \hline
leptons   &$\tau$   &1777[MeV] &4   &$\checkmark$   &$\checkmark$ \\ 
     &$\mu$   &106[MeV] &4   &$\checkmark$   &$\checkmark$ \\ 
     &$e$    &511[keV] &4   &$\checkmark$   &$\checkmark$ \\ 
     &$\nu_{\tau}$ &$\sim$0 &2   &$\checkmark$   &$\checkmark$ \\ 
     &$\nu_{\mu}$ &$\sim$0 &2   &$\checkmark$   &$\checkmark$ \\ 
     &$\nu_{e}$  &$\sim$0 &2   &$\checkmark$   &$\checkmark$ \\ \hline
gauge bosons &$W^{+}$   &80[GeV] &3   &$\checkmark$   &$\checkmark$ \\ 
     &$W^{-}$   &80[GeV] &3   &$\checkmark$   &$\checkmark$ \\ 
     &$Z$    &90[GeV] &3   &$\checkmark$   &$\checkmark$ \\ 
     &$\gamma$  &0    &2   &$\checkmark$   &$\checkmark$ \\ \hline
Higgs boson  &$H^{0}$   &125[GeV] &1   &$\checkmark$   &$\checkmark$ \\ \hline 
mesons   &$\pi^{0}$  &134[MeV] &1   &       &$\checkmark$ \\ 
     &$\pi^{+}$  &139[MeV] &1   &       &$\checkmark$ \\ 
     &$\pi^{-}$  &139[MeV] &1   &       &$\checkmark$ \\ \hline \hline
\end{tabular}
\caption{Input parameters of the relevant particles in the standard model before and after the QCD phase transition.}
\label{particlecontents}
\end{table}
From Fig.~\ref{scalaronmasshistory_fig1} 
one can see that the trace of energy-momentum tensor (black line) is always smaller than the energy density (red dashed line), implying that the pressure is certainly not negligible. 
In the low-temperature region, 
the trace of energy-momentum tensor damps quickly because the massless particles do not contribute although their energy densities remain.

\begin{figure}[htbp]
\centering
\includegraphics[width=0.6\textwidth]{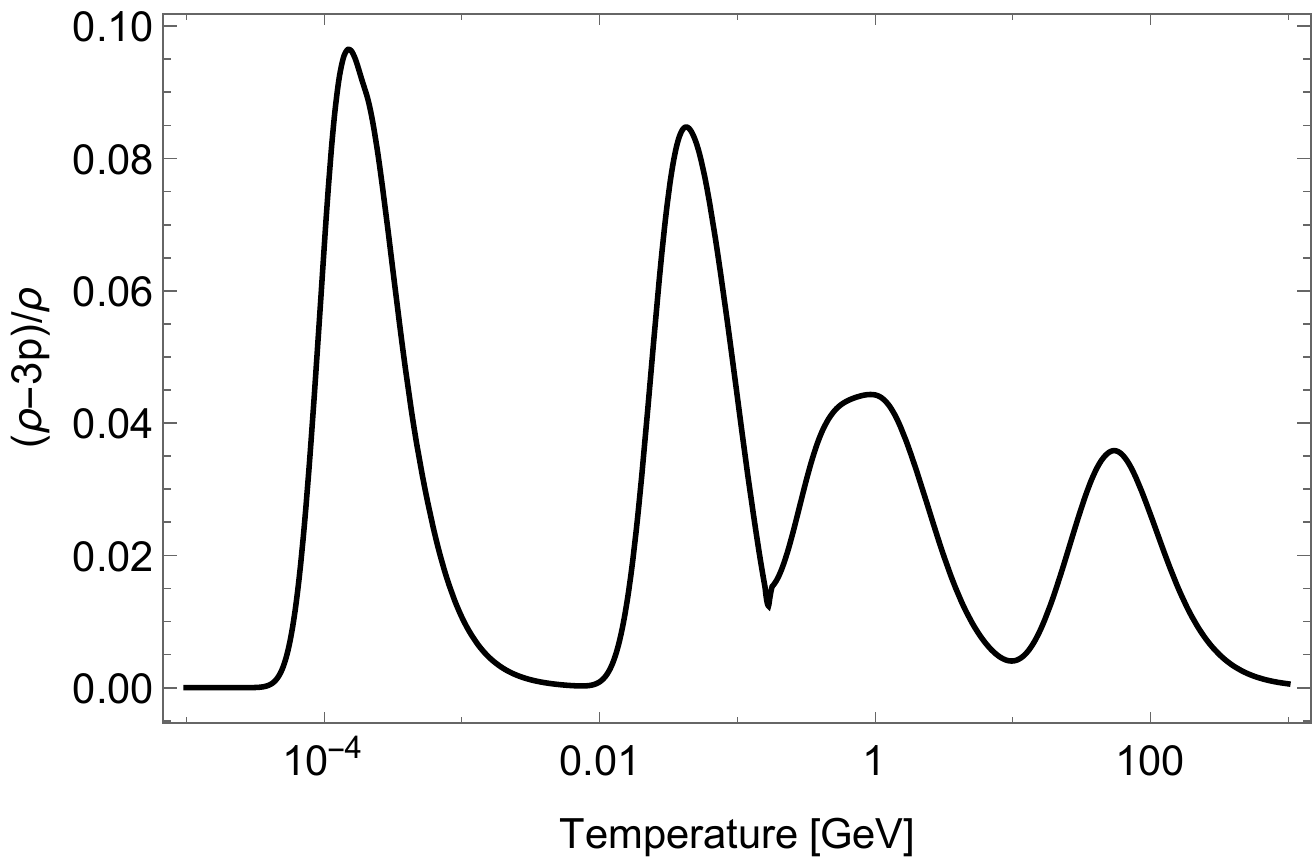}
\caption{The plot of $(\rho-3p)/\rho$ as a function of the temperature.}
\label{scalaronmasshistory_fig2}
\end{figure}

To compare $\rho - 3p$ with $\rho$, we evaluate the ratio between these two quantities, 
which is shown in Fig.~\ref{scalaronmasshistory_fig2}.
The peaks show up when the threshold condition $m \sim T$ is achieved. 
From the high-temperature side, 
the first peak corresponds to the net contribution of the threshold effects 
for top-quark, weak bosons, and Higgs;  
the second one for the charm and bottom quarks and tau; 
the third one below the QCD phase transition ($T=170$ [MeV]) 
for the mu and pions; 
the final peak in the low-temperature region for 
the electron. 

\begin{figure}[htbp]
\centering
\includegraphics[width=0.6\textwidth]{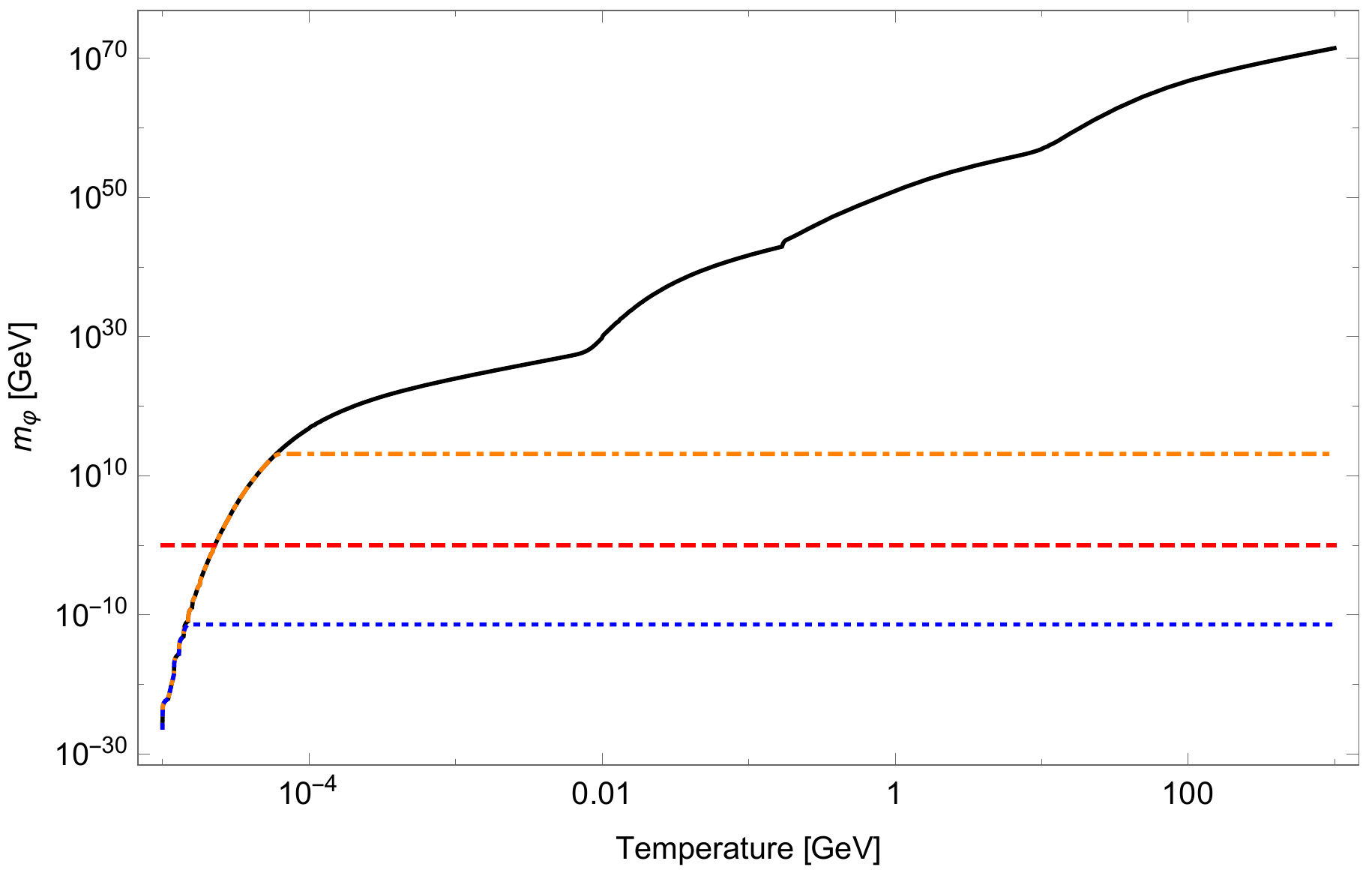}
\caption{The black curve represents the plot of the mass in the original Starobinsky model as a function of temperature,
with $\rho-3p$ in Fig.~\ref{scalaronmasshistory_fig1} substituted.
The orange dot-dashed and blue dotted lines respectively correspond to the masses with $\alpha = 10^{-27} [\mathrm{GeV}^{-2}]$ and $\alpha = 10^{22} [\mathrm{GeV}^{-2}]$ in the Starobinsky model with $R^{2}$ correction.
The red dashed line corresponds to $m_{\varphi} = 1 [\mathrm{GeV}] $, just for the reference value.}
\label{scalaronmasshistory_fig3}
\end{figure}

Finally, we study the time evolution of the scalaron mass 
by substituting the $(\rho-3 p)$ displayed in Fig.~\ref{scalaronmasshistory_fig2}  
into the mass formula Eq.~(\ref{scalaronmass2}). 
The result is shown in Fig.~\ref{scalaronmasshistory_fig3}.
The solid black line shows the scalaron mass in the original Starobinsky model, 
and dotted and dashed lines do in the Starobinsky model with $R^{2}$ correction.
The figure clearly shows that 
in the original Starobinsky model, 
the scalaron becomes extremely heavy 
and the mass becomes even heavier than the Planck mass $M_{\mathrm{pl}} \sim 10^{18} [\mathrm{GeV}]$. 
On the other hand, the scalaron mass is upper-bounded by including $R^2$ correction, 
and its upper limit is given as $m_{\varphi} \sim \alpha^{-1/2}$, as estimated in Eq.~(\ref{scalaronmass3}). 
In Fig.~\ref{scalaronmasshistory_fig3}, the orange dot-dashed line corresponds to
the case where the scalaron mass is on the order of the inflaton mass scale,
while the blue dotted line is obtained from the observational limit
\cite{Kapner:2006si,Adelberger:2006dh}.
Thus we find that the scalaron in the early Universe is characterized and controlled by the parameter $\alpha$, while
the scalaron mass becomes even less than $1 [\mathrm{GeV}]$ where the temperature is below $\mathcal{O}(10) [\mathrm{keV}]$, where the Big Bang nucleosynthesis happened. 
This result indeed supports our naive expectation on the possible cosmic history 
for the scalaron as noted in the previous subsection:  
{\it the scalaron can be heavy in the early Universe but should be light, at least, in the current Universe. }

\subsection{Scalaron production in thermal history}
\label{5D}

In the previous section, 
we have found that the scalaron mass behaves like almost constant in the early Universe, which strongly depends on the parameter $\alpha$ along with the $R^2$ correction. 
The value of the scalaron mass $m_\varphi$ (i.e., $\alpha$) is of importance when one considers the production mechanism of the scalaron in thermal history. 
It should also be noted that the scalaron interactions with the standard-model thermal bath are highly suppressed by the Planck scale (see Eqs.(\ref{Lag:FF})-(\ref{Lag:psi})), 
so that the scalaron interactions can never be thermalized.   
Nevertheless, the scalaron could non-thermally be produced as in the freeze-in scenarios~\cite{Blennow:2013jba} if the scalaron was moderately light in the early Universe.  
However, one can suspect that 
it is not an ordinary non-thermal production:  
because the scalaron dynamics have 
the close dependence on the thermal background made of 
the standard-model matters in the thermal history, 
one needs to solve the Boltzmann equation by taking into account 
the electroweak or QCD phase transitions, which would 
presumably be highly model-dependent and 
would practically be hard to address the estimate of the cosmic abundance.

Another possibility for the scalaron to accumulate the cosmic abundance is the non-thermal production via the coherent oscillation as in the case of axion-like dark matter. 
In that case, one needs to take the parameter $\alpha$ to be small enough 
(say $\alpha=10^{-27} [\mathrm{GeV}^{-2}]$ for $m_\varphi = 10^{13}$ [GeV] as in Fig.~\ref{scalaronmasshistory_fig3}),  
not to make the scalaron freeze-in in the early Universe. 
This possibility will be pursued closely in the next section. 
As it will turn out, 
this scenario is more intriguing than the freeze-in scenario, 
because one can address the coincidence problem between the dark energy and dark matter.  

Before proceeding the discussion on the coherent oscillation, 
we here make some comments: 
actually, the validity of the approximation as the harmonic oscillation is not easily 
verified because the scalaron potential minimum moves in time, although 
we can, in principle, follow up the all time-evolution dynamics of the scalaron once some particle production mechanism (i.e., inflation models) are fixed.  
Instead of addressing such specific models, 
to be more generic,  
in the present study, we will parameterize the scalaron 
amplitude during the harmonic oscillation by introducing 
some undetermined factor $(A)$, which actually controls 
the abundance today, in a way similar to the misalignment mechanism 
in the axion-like dark matter scenario. 
This uncertainty can be fed back to the model-dependence for inflation models addressing some particle production mechanism.  

\section{Relic abundance}
\label{5}

\subsection{Coincidence problem}
\label{5A}

In this subsection, we discuss the possibility to have the scalaron relic abundance today, 
which can be accumulated by the coherent oscillation mechanism. 
It is shown that the scalaron can naturally account for the cold dark matter abundance, as well as the dark energy density: the coincidence problem can be solved.

As long as the lifetime is longer than the age of Universe as in Eq.~(\ref{scalaronmassupperbound}), 
the scalaron is present today as a stable particle. 
The particle picture arises from 
fluctuating around the minimum of the potential $V(\varphi)$ 
by the shift $\varphi \to \varphi_{\min} + \varphi$, 
and its dynamics can be described by 
the equation of motion: 
\begin{align} 
\ddot{\varphi} + 3 H_{0} \varphi + \frac{\partial V(\varphi)}{\partial \varphi} 
= 0 
\,, 
\end{align}
with the dissipative term due to the Hubble parameter $H_{0}$ at present time.

We assume that the size of the fluctuation is sufficiently smaller than the Planck scale, $\varphi \ll M_{\rm pl}$. 
In that case the harmonic term dominates in 
the potential, so one can approximate the equation of motion 
to find the damping harmonic oscillation,   
\begin{align} 
\ddot{\varphi} + 3 H_{0} \varphi + \frac{1}{2} m_\varphi^2 \varphi^2 \approx 0 
\,. \label{damp:os:eq}
\end{align}
In terms of the scalaron energy density $\rho_\varphi 
= \frac{1}{2} \dot{\varphi}^2 + \frac{1}{2} m_\varphi^2 \varphi^2$, one can rewrite Eq.~(\ref{damp:os:eq}) 
to find that the $\rho_\varphi$ scales like nonrelativistic (pressure less) 
matter, $\rho_\varphi \sim a^{-3}$ with the scale factor $a$ related to the 
Hubble parameter $H$ as $H=(\dot{a}/a)$. In that sense, the scalaron at present day 
acts as a dark matter.

It turns out that the damping harmonic oscillation in Eq.~(\ref{damp:os:eq}) starts 
when the Hubble parameter $H$ reaches $\simeq m_\varphi/3$. 
During this oscillation, one finds an adiabatic invariant quantity $\propto (\rho_\varphi a^3)$, 
which can provide the currently observed cosmic abundance of the scalaron as dark matter.  
 For the scalaron to harmonically oscillate still at present, 
 the current scalaron mass should satisfy 
 $m_{\varphi} > 3 H_0$, which leads to the lower bound,  
\begin{align}
m_\varphi >6 \times 10^{-33} [\mathrm{eV} ]
\,, 
\label{harmonicoscillationmass1}
\end{align}
where we have used 
$H_{0}  \sim 2 \times 10^{-33} [\mathrm{eV}]$. 
Note that this limit is consistent with the constraint on the lifetime in Eq.~(\ref{scalaronmassupperbound}).

The energy density as the scalaron dark matter, averaged over the single period of the harmonic 
oscillation, is then evaluated as 
\begin{align}
(\rho_\varphi)_{DM} 
=\frac{1}{2} m^2_\varphi \varphi^2_0  
\label{scalaronrelicdensity1}
\,, 
\end{align}
with the amplitude $\varphi_0$. 
Combined with the potential energy at the minimum, which is nothing but the dark energy 
$(V(\varphi_{\min}) = (\rho_\varphi)_{DE})$, 
the total energy is then given by the sum of two: 
\begin{align}
(\rho_\varphi)_{\rm total} 
= (\rho_\varphi)_{DE} 
+ (\rho_\varphi)_{DM}
=  V(\varphi_{\min}) 
+ \frac{1}{2}m^2_\varphi \varphi_0^2
\,, 
\end{align} 
where 
\begin{align} 
(\rho_\varphi)_{DE} =   V(\varphi_{\min}) 
= M_\mathrm{pl}^{2} \Lambda
\,. 
\label{scalaronrelicdensity2}
\end{align}
One may assume that the 
scalaron fully explains the cold-dark matter component today, i.e. 
\begin{align} 
\rho_{CDM} 
= 
(\rho_\varphi)_{DM} 
\,, \label{CDM}
\end{align}
so as to take the numerical relation for observed amounts, 
$(\rho_\varphi)_{DM} \simeq 3/7 (\rho_\varphi)_{DE}$. 
Then one evaluates 
the amplitude of the harmonic oscillation, $\varphi_0$ in Eq.~(\ref{scalaronrelicdensity1}) 
as 
\begin{align}
\varphi^2_0 \simeq 
 \frac{6}{7} 
 M_{\rm pl}^2 \frac{\Lambda}{m^2_\varphi}
 \,.   
\label{scalaronrelicdensity4}
\end{align}
Furthermore, one may scale out 
the amplitude by the Planck mass scale $M_{\rm pl}$  
with a coefficient $A$ like 
$\varphi_0 =AM_{\rm pl}$, 
where we take $A\ll 1$ consistently with the 
approximation of the harmonic oscillation for the scalaron dynamics today.  
From Eq.~(\ref{scalaronrelicdensity4}) 
 the scalaron mass can then be expressed as 
\begin{align}
m^2_\varphi \simeq  \frac{6}{7} \frac{\Lambda}{A^{2}} \,  (\gg \Lambda) 
\,. 
\label{harmonicoscillationmass2}
\end{align}
By taking into account the harmonic oscillation condition in 
Eq.~(\ref{harmonicoscillationmass1}), 
the factor $A$ in Eq.~(\ref{harmonicoscillationmass2}) is 
now constrained to be 
 \begin{align}
A < 0.3 \, (\ll 1) 
\,, 
\end{align}
where we used $\Lambda \sim 4 \times 10^{-66} [\mathrm{eV^{2}}] $. 
This condition is indeed consistent with the harmonic oscillation approximation and can be rephrased as the constraint on some initial condition for particle production mechanism in modeling inflation scenarios.

Thus, the scalaron can naturally 
account for both the dark energy and dark matter without 
ad hoc tuning of parameters: 
{\it scalaron gives a possible solution to the coincidence problem!} 

\subsection{Scalaron in galaxies}
\label{5B}

\begin{figure}[htbp]
\centering
\includegraphics[width=0.6\textwidth]{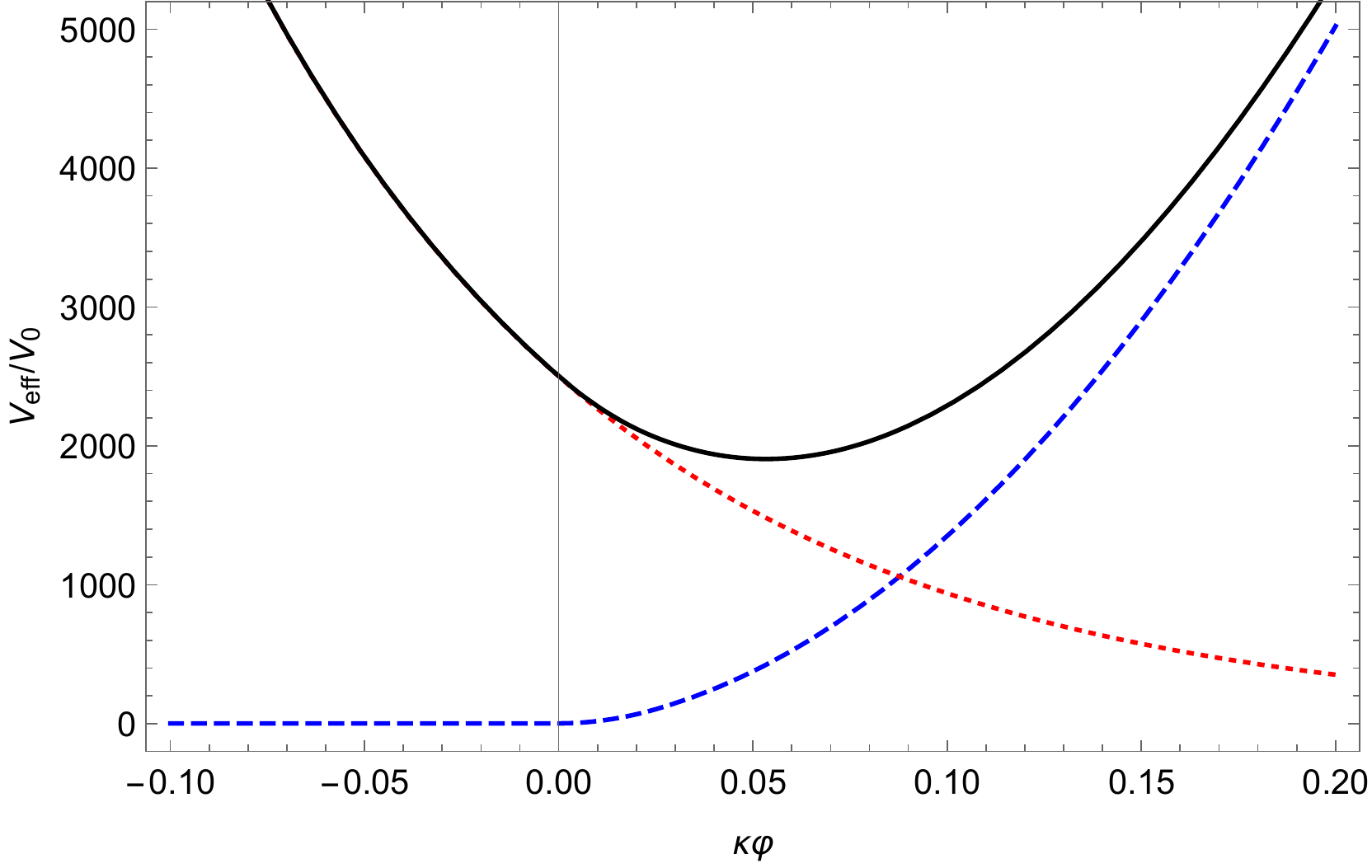}
\caption{The same as Fig.~\ref{scalaroneffectivepotential_fig2} 
for $- T^{\mu}_{\ \mu} \sim 10^{4} \cdot \Lambda/2\kappa^{2} \sim 10^{-25} [\mathrm{g/cm^{3}}]$.}
\label{scalaroneffectivepotential_fig9}
\end{figure}

Finally, in this subsection, we discuss the concrete example for the properties of the scalaron in the current Universe. 
As an example, we consider the situation in the galaxy
where the typical energy density of the galaxy under the dust approximation is given by 
$\rho \sim 3\mbox{--}5 \times 10^{-25} [\mathrm{g/cm^{3}}] \sim 2\mbox{--}3 \times 10^{-42}[\mathrm{GeV^{4}}]$ \cite{Sakamoto:2002zr}.
For this environment, the effective potential is given in Fig.~\ref{scalaroneffectivepotential_fig9}. 
Compared with Fig.~\ref{scalaroneffectivepotential_fig2},
one can find that 
the minimum of the potential moves toward the positive value, 
and that the curvature of the potential becomes large.
Actually, we find that the minimum is realized at $\kappa \varphi_{\min} \sim 0.05$,
and the scalaron mass is computed as 
\begin{align} 
m_{\varphi} \sim 3\mbox{--}5 \times 10^{-24}  [\mathrm{eV}]
\,. \label{mass:galaxy}
\end{align} 
We now discuss the scalaron mass in the galaxy.
Apparently, the scalaron mass in Eq.(\ref{mass:galaxy}) 
is very small and naturally satisfies the mass bound $m_{\varphi} < \mathcal{O}(1) [\mathrm{GeV}]$ in Eq.~(\ref{scalaronmassupperbound}).
It is remarkable that the estimated size of mass in Eq.(\ref{mass:galaxy})
is so close to the value $m \sim 10^{-23}  [\mathrm{eV}]$ for the ultralight axion which solves the core cusp problem of the galaxies \cite{Hu:2000ke}. 
The bound for the axion as a self-interacting dark matter 
would be applicable also to the scalaron. .
It would be a salient result that the ultralight ``scalaron" possibly solves the cusp problem in the galaxy, 
although the input parameter in $F(R)$ gravity model has nothing to do with this problem.

\section{Conclusion and Discussion}

In this paper,
we have discussed the new scalar field, the scalaron, introduced from the modification of gravity through the Weyl transformation.
We have assumed that 
the oscillation around the potential minimum of the scalaron can be interpreted as the dark matter, while the potential energy gives the dark energy.

In the first part of this paper,
we revisited our previous work and refined the upper bound for the scalaron mass.
We assumed that the decay process of the scalaron to two photons is dominated at the late-time,
and we obtained the upper bound for the scalaron mass $m_{\varphi} < \mathcal{O}(1) [\mathrm{GeV}]$.
We also discussed the scalaron mass in the Starobinsky model with $R^{2}$ correction.
In our previous work, we had found that 
the scalaron mass becomes too large in the original Starobinsky model, even larger than the Planck Mass.
And then, the parameter in the Starobinsky model was tightly constrained,
and such a parameter region was irrelevant to the dark energy problem.
We found that the above problem can be related to the singularity problem in $F(R)$ gravity,
and added the $R^{2}$ term in the action to cure this problem.
We also found that the scalaron mass is controllable to be smaller than the Planck scale, 
due to the $R^{2}$ term.

In the second part, 
we evaluated the time-evolution of the scalaron mass in the Universe,
especially, in the early and current Universe.
To take into account the chameleon mechanism, 
we have constructed the energy-momentum tensor composed by the standard model particles to reproduce the environment in the early Universe.
By using the time-evolution of the trace of energy-momentum tensor,
we discussed the cosmic history of the scalaron mass, 
and confirmed that the scalaron mass becomes smaller and smaller according to the chameleon mechanism.
We addressed possible production mechanisms of the scalaron in the thermal history,
and especially studied the non-thermal production.
Then, we discussed the relic abundance as dark matter.
We found that the relic energy density as the dark matter is naturally comparable to that as the dark energy if the approximation of the harmonic oscillation around the potential minimum is valid.
Finally, we studied the concrete example of the environment in the current Universe.
We calculated the scalaron mass in the galaxy, 
and then found that the scalaron is very light $m_{\varphi} \sim 10^{-24}  [\mathrm{eV}]$,
comparable to the ultralight axion dark matter.
Thus, we can conclude that the $F(R)$ gravity can explain the dark matter and dark energy.

In closing, we give several comments.
We have shown that the coincidence problem would be addressed
as long as the harmonic oscillation approximation of the scalaron is valid.
However, we did not predict the numerical value of the energy density as dark matter,
rather input the observed value.
It is possible in principle, but technically difficult to calculate the amplitude $A$ of the oscillation in the current Universe
because we need to reproduce the time evolution of comic environment.
If we should calculate the time evolution of the scalaron potential, 
we could compute the amplitude of the oscillation and evaluate the relic abundance of the scalaron dark matter.

Regarding the validity of the harmonic oscillation of the scalaron,
the contribution from the standard model particles to the effective potential is important so as to make the potential lifted up to have the minimum.
This is closely related to the particle production scenario.
If we additionally introduce the inflaton field for the inflation,
the scalaron couples with the inflaton, and the scalaron potential would be lifted up as desired.
It is even interesting to study the case that the $R^{2}$ correction can be regarded as $R^{2}$ inflation.
In that case, we could unify all cosmic history in terms of the $F(R)$ gravity.
However, we need to discuss the compatibility with the particle production.
In the standard particle production scenario,
the inflaton decays and produces the standard model particles
during the oscillation at the potential minimum.
In contrast, the scalaron potential does not have the minimum without matters in the Starobinsky model with $R^{2}$ correction.
Thus, the naive reheating scenario is not applicable if the scalaron also plays a role of the inflaton.
We then need to invoke the preheating scenario or beyond-standard-model particles.

As to the phenomenology of the scalaron dark matter,
it is necessary to distinguish the scalaron from other dark matter candidates,
especially, axion-like particles. 
Of importance is to notice that the scalaron has the chameleon mechanism.
For example, 
the scalaron would be heavy in the atmosphere of the Earth
although the scalaron is light in the interstellar environment as we discussed in this paper.
Therefore, such a chameleonic particle would be observed by analyzing the difference of the experimental data from the different experimental environments.
Tests for the chameleon mechanism also gives us the constraint on the scalaron dark matter,
and then, we possibly obtain the two independent ways to search the scalaron dark matter.
About indirect detection,
heavy scalaron could be observed in the decay to the photons.
If it occurs at galactic center where the energy density is very large,
the heavy scalaron decays to energetic photons,
and emitted photons should be observed in the cosmic ray.
It is also quite interesting to apply the chameleonic particle for other fields of particles physics, as discussed in \cite{Nelson:2008tn}.

\section*{Acknowledgments}
Authors would like to give special thanks to Shin'ichi Nojiri and Junji Hisano for the fruitful discussions,
and to Masaharu Tanabashi for the useful comments.
This work is supported by Grant-in-Aid for Scientific research from
the Ministry of Education, Science, Sports, and Culture (MEXT), Japan, No. 16H06492 (T.K.),
by the JSPS Grant-in-Aid for Young Scientists (B) No. 15K17645 (S.M.).

\appendix

\section{Calculation of trace of energy-momentum tensor}
\label{A1}
In this appendix, we show the detailed calculation to derive the analytic form of  $\rho - 3p$.
At the high temperature, we take the limit $x \ll 1$, 
and then the integral in Eq.~(\ref{emttrace1}) is approximated as
\begin{align}
\int^{\infty}_{0} dy \frac{y^{2}}{\sqrt{x^{2} + y^{2}}} \frac{1}{\mathrm{e}^{\sqrt{x^{2} + y^{2}}} \pm 1} 
\approx &
\int^{\infty}_{0} dy \frac{y}{\mathrm{e}^{y} \pm 1} 
\, .
\end{align}
For bosons, we use the following formula:
\begin{align}
\int^{\infty}_{0} dy \frac{y^{n}}{\mathrm{e}^{y} - 1} 
=&
\zeta(n+1) \Gamma (n+1)
\, .
\end{align}
Thus, the integral is given by
\begin{align}
\int^{\infty}_{0} dy \frac{y}{\mathrm{e}^{y} - 1} 
=&
2 \zeta(2)
\, ,
\end{align}
where $\zeta(2) = \frac{\pi^{2}}{6}$.
For fermions, we have
\begin{align}
\int^{\infty}_{0} dy \frac{y}{\mathrm{e}^{y} + 1} 
=&
\int^{\infty}_{0} dy 
\left[
\frac{y}{\mathrm{e}^{y} - 1} - \frac{2y}{\mathrm{e}^{2y} - 1}
\right]
\nonumber \\
=&
\zeta(2)
\, .
\end{align}
Hence, we obtain the following expression in the relativistic limit:
\begin{align}
\rho - 3p \approx \frac{g}{12}m^2T^{2}
\left\{
\begin{array}{l}
2\ \mbox{for bosons} \\
1\ \mbox{for fermions}
\end{array}
\right.
\, .
\end{align}

At the low temperature, we take the limit $x \gg 1$, 
and then the integral in Eq.~(\ref{emttrace1}) is approximated as
\begin{align}
&\int^{\infty}_{0} dy \frac{y^{2}}{\sqrt{x^{2} + y^{2}}} \frac{1}{\mathrm{e}^{\sqrt{x^{2} + y^{2}}} \pm 1} 
\nonumber \\
&\approx
\int^{\infty}_{0} dy \frac{y^{2}}{\sqrt{x^{2} + y^{2}}} \frac{1}{\mathrm{e}^{\sqrt{x^{2} + y^{2}}} } 
\nonumber \\
&\approx
\frac{1}{x} \mathrm{e}^{-x} \int^{\infty}_{0} dy y^{2} \left( 1 - \frac{y^{2}}{2x^{2}} \right) \mathrm{e}^{- \frac{y^{2}}{2x}}  
\nonumber \\
&=
\frac{1}{x} \mathrm{e}^{-x} \int^{\infty}_{0} dy y^{2} \mathrm{e}^{- \frac{y^{2}}{2x}}  
-  \frac{1}{2x^{3}} \mathrm{e}^{-x} \int^{\infty}_{0} dy y^{4} \mathrm{e}^{- \frac{y^{2}}{2x}}  
\, .
\end{align}
Here, we use the following formula:
\begin{align}
\int^{\infty}_{0} dy y^{n} \mathrm{e}^{-y^{2}}
=& 
\frac{1}{2} \Gamma \left( \frac{1}{2}(n+1) \right)
\, .
\end{align}
Thus, the integral is given by
\begin{align}
&\int^{\infty}_{0} dy \frac{y^{2}}{\sqrt{x^{2} + y^{2}}} \frac{1}{\mathrm{e}^{\sqrt{x^{2} + y^{2}}} \pm 1} 
\nonumber \\
&\approx
\frac{1}{x} \mathrm{e}^{-x} \cdot \frac{1}{2} \Gamma \left( \frac{3}{2} \right) (2x)^{3/2}
-  \frac{1}{2x^{3}} \mathrm{e}^{-x} \cdot \frac{1}{2} \Gamma \left( \frac{5}{2} \right) (2x)^{5/2}
\nonumber \\
&=
\sqrt{ \frac{\pi}{2} } x^{1/2} \mathrm{e}^{-x} \left(1  - \frac{3}{2}  x^{-1}  \right)
\, .
\end{align}
We then obtain the following expression in the non-relativistic limit:
\begin{align}
\rho - 3p 
=& 
\frac{g T^{4}}{2 \pi^{2}} \cdot x^{2}  
\sqrt{ \frac{\pi}{2} } x^{1/2} \mathrm{e}^{-x} \left(1  - \frac{3}{2}  x^{-1}  \right)
\nonumber \\
=&
m g \left( \frac{m T}{2 \pi} \right)^{3/2} \mathrm{e}^{-\frac{m}{T}}   - \frac{3}{2} g T \left( \frac{m T}{2 \pi} \right)^{3/2} \mathrm{e}^{-\frac{m}{T}} 
\, .
\end{align}


\begin{thebibliography}{99}


\bibitem{Starobinsky:1980te} 
  A.~A.~Starobinsky,
  Phys.\ Lett.\  {\bf 91B}, 99 (1980).
  doi:10.1016/0370-2693(80)90670-X


\bibitem{Nojiri:2017ncd} 
  S.~Nojiri, S.~D.~Odintsov and V.~K.~Oikonomou,
  Phys.\ Rept.\  {\bf 692}, 1 (2017)
  doi:10.1016/j.physrep.2017.06.001
  [arXiv:1705.11098 [gr-qc]].


\bibitem{Katsuragawa:2016yir} 
  T.~Katsuragawa and S.~Matsuzaki,
  Phys.\ Rev.\ D {\bf 95}, no. 4, 044040 (2017)
  doi:10.1103/PhysRevD.95.044040
  [arXiv:1610.01016 [gr-qc]].

\bibitem{Nojiri:2008nk} 
  S.~Nojiri and S.~D.~Odintsov,
  arXiv:0801.4843 [astro-ph].

\bibitem{Nojiri:2008nt} 
  S.~Nojiri and S.~D.~Odintsov,
  TSPU Bulletin N {\bf 8(110)}, 7 (2011)
  [arXiv:0807.0685 [hep-th]].

\bibitem{Cembranos:2008gj} 
  J.~A.~R.~Cembranos,
  Phys.\ Rev.\ Lett.\  {\bf 102}, 141301 (2009)
  doi:10.1103/PhysRevLett.102.141301
  [arXiv:0809.1653 [hep-ph]].

\bibitem{Choudhury:2015zlc} 
  S.~Choudhury, M.~Sen and S.~Sadhukhan,
  Eur.\ Phys.\ J.\ C {\bf 76}, no. 9, 494 (2016)
  doi:10.1140/epjc/s10052-016-4323-2
  [arXiv:1512.08176 [hep-ph]].





\bibitem{Starobinsky:2007hu} 
  A.~A.~Starobinsky,
  JETP Lett.\  {\bf 86}, 157 (2007)
  doi:10.1134/S0021364007150027
  [arXiv:0706.2041 [astro-ph]].


\bibitem{Khoury:2003aq} 
  J.~Khoury and A.~Weltman,
  Phys.\ Rev.\ Lett.\  {\bf 93}, 171104 (2004)
  doi:10.1103/PhysRevLett.93.171104
  [astro-ph/0309300].


\bibitem{Frolov:2008uf} 
  A.~V.~Frolov,
  Phys.\ Rev.\ Lett.\  {\bf 101}, 061103 (2008)
  doi:10.1103/PhysRevLett.101.061103
  [arXiv:0803.2500 [astro-ph]].

\bibitem{Nojiri:2008fk} 
  S.~Nojiri and S.~D.~Odintsov,
  Phys.\ Rev.\ D {\bf 78}, 046006 (2008)
  doi:10.1103/PhysRevD.78.046006
  [arXiv:0804.3519 [hep-th]].

\bibitem{Dev:2008rx} 
  A.~Dev, D.~Jain, S.~Jhingan, S.~Nojiri, M.~Sami and I.~Thongkool,
  Phys.\ Rev.\ D {\bf 78}, 083515 (2008)
  doi:10.1103/PhysRevD.78.083515
  [arXiv:0807.3445 [hep-th]].

\bibitem{Bamba:2008ut} 
  K.~Bamba, S.~Nojiri and S.~D.~Odintsov,
  JCAP {\bf 0810}, 045 (2008)
  doi:10.1088/1475-7516/2008/10/045
  [arXiv:0807.2575 [hep-th]].

\bibitem{Kobayashi:2008wc} 
  T.~Kobayashi and K.~i.~Maeda,
  Phys.\ Rev.\ D {\bf 79}, 024009 (2009)
  doi:10.1103/PhysRevD.79.024009
  [arXiv:0810.5664 [astro-ph]].

\bibitem{Capozziello:2009hc} 
  S.~Capozziello, M.~De Laurentis, S.~Nojiri and S.~D.~Odintsov,
  Phys.\ Rev.\ D {\bf 79}, 124007 (2009)
  doi:10.1103/PhysRevD.79.124007
  [arXiv:0903.2753 [hep-th]].



\bibitem{Brax:2004qh} 
  P.~Brax, C.~van de Bruck, A.~C.~Davis, J.~Khoury and A.~Weltman,
  Phys.\ Rev.\ D {\bf 70}, 123518 (2004)
  doi:10.1103/PhysRevD.70.123518
  [astro-ph/0408415].

\bibitem{Husdal:2016haj} 
  L.~Husdal,
  Galaxies {\bf 4}, no. 4, 78 (2016)
  doi:10.3390/galaxies4040078
  [arXiv:1609.04979 [astro-ph.CO]].


\bibitem{Kapner:2006si} 
  D.~J.~Kapner, T.~S.~Cook, E.~G.~Adelberger, J.~H.~Gundlach, B.~R.~Heckel, C.~D.~Hoyle and H.~E.~Swanson,
  Phys.\ Rev.\ Lett.\  {\bf 98}, 021101 (2007)
  doi:10.1103/PhysRevLett.98.021101
  [hep-ph/0611184].

\bibitem{Adelberger:2006dh} 
  E.~G.~Adelberger, B.~R.~Heckel, S.~A.~Hoedl, C.~D.~Hoyle, D.~J.~Kapner and A.~Upadhye,
  Phys.\ Rev.\ Lett.\  {\bf 98}, 131104 (2007)
  doi:10.1103/PhysRevLett.98.131104
  [hep-ph/0611223].




\bibitem{Blennow:2013jba}
  M.~Blennow, E.~Fernandez-Martinez and B.~Zaldivar,
  JCAP {\bf 1401} (2014) 003
  doi:10.1088/1475-7516/2014/01/003
  [arXiv:1309.7348 [hep-ph]]; 
  L.~J.~Hall, K.~Jedamzik, J.~March-Russell and S.~M.~West,
  JHEP {\bf 1003} (2010) 080
  doi:10.1007/JHEP03(2010)080
  [arXiv:0911.1120 [hep-ph]].


\bibitem{Sakamoto:2002zr} 
  T.~Sakamoto, M.~Chiba and T.~C.~Beers,
  Astron.\ Astrophys.\  {\bf 397}, 899 (2003)
  doi:10.1051/0004-6361:20021499
  [astro-ph/0210508].

\bibitem{Hu:2000ke} 
  W.~Hu, R.~Barkana and A.~Gruzinov,
  Phys.\ Rev.\ Lett.\  {\bf 85}, 1158 (2000)
  doi:10.1103/PhysRevLett.85.1158
  [astro-ph/0003365].



\bibitem{Nelson:2008tn} 
  A.~E.~Nelson and J.~Walsh,
  Phys.\ Rev.\ D {\bf 77}, 095006 (2008)
  doi:10.1103/PhysRevD.77.095006
  [arXiv:0802.0762 [hep-ph]].


\end{thebibliography}
\end{document}